\documentclass[10pt,aps,prd,reprint,nofootinbib,floats,floatfix,amsfonts,amssymb,amsmath,preprintnumbers,notitlepage,superscriptaddress]{revtex4-1}

\usepackage[utf8]{inputenc}
\usepackage[margin=1in]{geometry}
\usepackage{graphicx}
\usepackage{xcolor}
\usepackage[colorlinks]{hyperref}
\usepackage{amsmath, amssymb, amsfonts,amsbsy,amsthm} 
\usepackage[normalem]{ulem}

\newcommand{\Kappa}[0]{\scalebox{1.5}{$\kappa$}}
\newcommand{\orcid}[1]{\href{https://orcid.org/#1}{\includegraphics[scale=0.15]{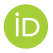}}}

\begin{document}

\title{Correlated 1-1000 Hz magnetic field fluctuations from lightning over earth-scale distances and their impact on gravitational wave searches}

\author{Kamiel Janssens\orcid{0000-0001-8760-4429}}
\affiliation{Universiteit Antwerpen, Prinsstraat 13, 2000 Antwerpen, Belgium}
\affiliation{Universit\'e C$\hat{o}$te d’Azur, Observatoire C$\hat{o}$te d’Azur, CNRS, ARTEMIS, Nice, France}

\author{Matthew Ball\orcid{0000-0001-5565-8027}}
\affiliation{University of Oregon, Eugene, OR 97403, USA}

 \author{Robert M. S. Schofield}
 \affiliation{University of Oregon, Eugene, OR 97403, USA}
 
\author{Nelson Christensen\orcid{0000-0002-6870-4202}}
\affiliation{Universit\'e C$\hat{o}$te d’Azur, Observatoire C$\hat{o}$te d’Azur, CNRS, ARTEMIS, Nice, France}

\author{Raymond Frey\orcid{0000-0003-0341-2636}}
\affiliation{University of Oregon, Eugene, OR 97403, USA}

\author{Nick van Remortel\orcid{0000-0003-4180-8199}}
\affiliation{Universiteit Antwerpen, Prinsstraat 13, 2000 Antwerpen, Belgium}

\author{Sharan Banagiri\orcid{0000-0001-7852-7484}}
\affiliation{School of Physics and Astronomy, University of Minnesota, Minneapolis, Minnesota 55455, USA}
\affiliation{Center for Interdisciplinary Exploration and Research in Astrophysics (CIERA), Northwestern University, 1800 Sherman Ave, Evanston, IL 60201, USA }

\author{Michael W. Coughlin\orcid{0000-0002-8262-2924}}
\affiliation{School of Physics and Astronomy, University of Minnesota, Minneapolis, Minnesota 55455, USA}

\author{Anamaria Effler\orcid{0000-0001-8242-3944}}
\affiliation{LIGO Livingston Observatory, Livingston, Louisiana 70754, USA}

\author{Mark Go{\l}kowski}
\affiliation{{University of Colorado Denver, Denver, CO 80204, USA}}

\author{Jerzy Kubisz}
\affiliation{Astronomical Observatory, Jagiellonian University, Krakow 30-244, Poland}

\author{Micha{\l} Ostrowski}
\affiliation{Astronomical Observatory, Jagiellonian University, Krakow 30-244, Poland}

\date{\today}

\begin{abstract}
We report Earth-scale distance magnetic correlations from lightning strokes in the frequency range 1--1000\,Hz at several distances ranging from 1100 to 9000\,km. 
Noise sources which are correlated on Earth-scale distances can affect future searches for gravitational-wave signals with ground-based gravitational-wave interferometric detectors. 
We consider the impact of correlations from magnetic field fluctuations on gravitational-wave searches due to Schumann resonances (<50 Hz) as well as higher frequencies (>100 Hz).
We demonstrate that individual lightning strokes are a likely source for the observed correlations in the magnetic field fluctuations at gravitational-wave observatories and discuss some of their characteristics. Furthermore, we predict their impact on searches for an isotropic gravitational-wave background, as well as for searches looking for short-duration transient gravitational waves, both unmodeled signals (bursts) as well as modeled signals (compact binary coalescence). Whereas the recent third observing run by LIGO and Virgo was free of an impact from correlated magnetic field fluctuations, future runs could be affected. 
For example, at current magnetic coupling levels,  neutron star inspirals in third generation detectors are likely to be contaminated by multiple correlated lightning glitches.
We suggest that future detector design should consider reducing lightning coupling by, for example, reducing the lightning-induced beam tube currents that pass through sensitive magnetic coupling regions in current detectors. We also suggest that the diurnal and seasonal variation in lightning activity may be useful in discriminating between detector correlations that are produced by gravitational waves and those produced by lightning. 
\end{abstract}

\maketitle


\section{Introduction}
\label{sec:Introduction}

Within the gravitational wave (GW) community, it is known that Schumann resonances  \cite{Schumann1,Schumann2} can impact searches for the gravitational-wave background (GWB) \cite{SGWBRevChristensen,Thrane:2013npa,Thrane:2014yza,10.1088/0264-9381/33/22/224003,PhysRevD.97.102007,Himemoto:2017gnw,Himemoto:2019iwd}. The Schumann resonances are electromagnetic excitations in the cavity produced by the ionosphere and the Earth's surface, sourced by lightning strikes. 
Short duration, magnetic transient signals have also been observed in coincidence on Earth-scale distances, and their implications for GW searches have been studied~\cite{10.1088/1361-6382/aa60eb}. The magnetic component from the Schumann resonances can couple to GW detectors in multiple ways, e.g. by direct coupling to permanent magnets in actuation systems or by acting upon electronics and cabling \cite{Thrane:2013npa,Thrane:2014yza,galaxies8040082,Nguyen_2021}. Before the most recent third observing run (O3), Advanced LIGO \cite{AdvancedLIGO} and Advanced Virgo \cite{AdvancedVirgo} installed low-noise magnetometers -- LEMI-120 at LIGO Hanford (H or LHO) and LIGO Livingston (L or LLO), Metronix MFS-06e at Virgo (V) -- in quiet locations, generally unaffected by local magnetic noise, at least few hundred meters away from any building, to monitor magnetic signals with high precision, such as the Schumann resonances \citep{PhysRevD.97.102007}. This is the first time these low noise magnetic measurements were recorded on a time scale of about one year at the LIGO and Virgo sites.
In this work we will not discuss measurements by the low-noise magnetometers installed at KAGRA\cite{PhysRevD.88.043007}.  

In this paper, we will study Earth-scale distance correlations of magnetic field fluctuations in the frequency range of 1--1000\,Hz, using a global network of low noise magnetometers located at GW detectors as well as low noise magnetometers located in the Bieszczady Mountains in Poland \cite{PolishMag}, which are part of the WERA project \cite{WERAProj}. The observed correlations will be discussed as a function of their separation (1100--9000\,km) as well as the characteristics and geography of lightning strokes\footnote{The term ``strike'' is colloquially used for lightning. However, in the technical literature, ``stroke'' describes both cloud-to-cloud and cloud-to-ground lightning. ``Strike'' only refers to cloud-to-ground. Later in this paper we will use the Vaisala GLD360 database \cite{Said2013} which detects ``strokes'' in general, so we use this term to refer to all detected lightning events here~\cite{strokeHolle}.} and thunderstorms. 

In Sec. \ref{sec:Geophysics} we focus on the geophysical processes of lightning strokes and the propagation of electromagnetic fields in the Earth's atmosphere.
In Sec. \ref{sec:IndivLightnings} we will investigate the coherence and correlations from the superposition of individual lightning strokes. In the rest of the paper we will often use `\textit{magnetic coherence}'  and `\textit{magnetic correlations}' to refer to the coherence and correlations, respectively, between the magnetic field fluctuations.
In Sec. \ref{sec:StochPoV} we will use the knowledge gained from Sec. \ref{sec:Geophysics} and Sec. \ref{sec:IndivLightnings} to compare the observed correlations between different sensor pairs.
The impact of these magnetic correlations will be discussed in Sec. \ref{sec:ImpactfutureSearches} where we will start by discussing how magnetic fields can couple to GW interferometric detectors (Sec. \ref{sec:magneticcoupling}). Afterwards we make a prediction of magnetic contamination with respect to a GWB in the frequency range 20Hz-675Hz (Sec. \ref{sec:impactGWB}). Furthermore, we will also predict the effect from the individual lightning strokes on searches for short duration GW transients, both unmodeled bursts and modeled compact binary coalescence (CBC) events (Sec. \ref{sec:impactBurst}).


\section{Properties of lightning strokes}
\label{sec:Geophysics}

In this section, we will discuss some characteristic properties of where lightning strokes mainly take place as well as how their signals propagate.

Long distance propagation of lightning induced impulsive radio waves (often called ``sferics'' in the literature) is affected by the radiated spectral content of the lightning discharge and also  the properties of the propagation channel in the Earth-ionosphere waveguide. The lightning source  is often quantified in terms of the peak current ($\sim$10$-$500 kA) and the charge moment ($\sim$25$-$ 500 C km). The lightning spectrum peaks in the central Very Low Frequency (VLF) band (10\,kHz$-$20\,kHz) but is very broadband \cite{cotts2011ionospheric}.   A larger charge moment is indicative of continuing currents and as such is correlated with higher Extremely Low Frequency (ELF) (3\,Hz$-$3\,kHz) spectral content. There is  evidence of a regional dependence to lightning properties with lightning over the oceans   characterized by large peak currents even though the majority of lightning events are over land and in coastal areas \cite{Said2013}. In some works, large charge moments have been associated with the African continent \cite{golkowski2011estimation, chen2008global}. The propagation path dependencies include the conductivity of the electron density profile of the lower ionosphere, the conductivity of the ground, crossing of the day-night terminator, and the azimuthal direction of propagation relative to the magnetic dip angle for low latitudes.  For the VLF band the overall daytime and nighttime attenuation rates are in the range  2–5 dB/1000 km and 1–3 dB /1000 km, respectively \cite{golkowski2011estimation, wait1957attenuation}. For the ELF band, attenuation increases with frequency and below 100 Hz can be 1 dB/1000 km or less\cite{ginsberg1974extremely}.  Attenuation for a typical day-night transition\footnote{The propagating wave makes the transition from the side of the Earth facing the sun to the the side not facing the sun.} is 1 dB and up to 3 dB for a night-day transition  \cite{wait1981lectures}. For low latitudes, attenuation for westward propagating waves can be 1.9$-$2.0 dB/1000 km higher as compared to eastward propagating waves for both daytime and nighttime ionospheric conditions \cite{hutchins2013azimuthal}. Propagation over low ground conductivity can add several dB/1000 km of attenuation as compared to propagation over conductive sea water.

There are three main global thunderstorm centers where most lightning events occur.  These are the equatorial regions of the Americas, sub-Saharan Africa and southeast Asia \cite{boccippio2000regional}.  Lightning strokes are most prevalent in the late afternoon with a minimum in late morning. In geographical regions dominated by large mesoscale convective systems, like the central United States, the peak in lightning prevalence shifts to the late evening or early morning hours. Lightning activity is maximum in June–August, corresponding to the Northern Hemisphere summer, while the minimum occurs in December–February~\cite{blakeslee2014seasonal}.


\section{Magnetic coherence between LIGO sites produced by the superposition of individual lightning strokes}
\label{sec:IndivLightnings}

\subsection{Establishing LIGO detections of individual lightning strokes}
\label{sec:establishLightnings}

Electromagnetic waves created by individual lightning strokes have signatures in both ELF and VLF bands \cite{Simoes2012}. Traditionally, the VLF band is used by lightning detection networks to track individual lightning signals, but the ELF band has also been shown to be usable for this purpose \cite{Simoes2012} and has some unique propagation characteristics \cite{golkowski2018ionospheric}. The ability of the ELF band to travel long distances implies that these signals could be picked up by sensitive magnetometers at GW detectors, even if the lightning occurs several hundreds to thousands kilometers away. This naturally raises the question whether the global lightning background could impact GW detectors. 

In the remainder of this section, we will focus
on coherence between LHO and LLO and we will discuss other sites in section \ref{sec:StochPoV}. 

We establish these correlated short-duration signals by identifying magnetic signals at individual magnetometers at each LIGO site. We then reduce these to magnetic signals coincident between sites to build a set of LIGO-measured magnetic signals. We then compare these to the GLD360 lightning dataset from Vaisala \cite{Said2013} as a confirmed reference. GLD360 data for the week of September 23, 2019 was provided for this section.

To identify lightning signals coincident between sites, we use the Omicron transient detection tool \cite{ROBINET2020100620} to select periods of excess noise in the LEMI-120 magnetometers located at LHO and LLO during the week of September 23, 2019. The LEMI-120 magnetometers are low-noise, sensitive up to around 400 \,Hz, and sampled at a rate of 4096\,Hz. Each site has two of these magnetometers, one aligned with each orthogonal arm of the interferometer, and the Omicron tool was run on each magnetometer independently. The Omicron tool looks for excess energy (deviation from the average background) via sine-Gaussian fits. This gives a time-frequency grid of energy for the data. SNR values are computed for each pixel of the grid, and those exceeding a minimum threshold are recorded. Individual pixels adjacent in time are clustered as combined events. Omicron returns the start and end time of each of these events (referred to as triggers) with the minimum and maximum frequency of the pixel cluster. It also returns the time and frequency of the maximum energy pixel of the trigger which are recorded as ``peak times'' and ``peak frequencies'' \cite{ROBINET2020100620}.

A comprehensive list of coincident LEMI signals was collected by selecting Omicron triggers meeting time coincidence criteria. For triggers in different magnetometers at the same site, we want trigger times to be effectively identical. For triggers at different sites, we want triggers to be separated by no more than the travel time for a wave propagating at $75 \%$ of the speed of light along the surface of the Earth. This accounts for the frequency-dependence of the signal propagation that decreases to as low as $0.75 c$ at $\sim$10Hz \cite{KEMP1971919}.
To account for uncertainties in Omicron time clustering, these thresholds are loosened to 5\,ms (20 LEMI samples) between magnetometers at the same site and twice the light travel time for magnetometers at different sites. 
This large uncertainty on the timing of the lightning is caused by the fact that Omicron combines both time and frequency information. Whereas the frequency information can give more insight on the spectral behaviour of the event it comes at a cost with respect to the time resolution.
Because each site has two orthogonal magnetometers, we selected triggers that were detected in at least three of the four total magnetometers subject to the aforementioned time delays. The peak time at each site was recorded for each coincident trigger to preserve the relative time delay between sites. This set of triggers, those coincident between LHO and LLO magnetometers, we refer to as ``HL coincident triggers.''

We match the HL coincident triggers to the Vaisala GLD360 lightning detection database \cite{Said2013}. This dataset is collected by a broad network of GPS-synchronized sensors which use matched waveforms to estimate the location of a lightning stroke as well as the peak current. The current is estimated with a resolution of $\mathcal{O}$(kA), with polarity determined by the direction of the electric field -- a negative current is assigned for an electric field pointing toward the Earth \cite{Said2013}. The GLD360 network claims an $80\%$ detection efficiency in identifying cloud-to-ground strokes in the Northern Hemisphere and between $10\%$ and $80\%$ detection efficiency in the Southern Hemisphere \cite{vaisala_efficiency_1,vaisala_efficiency_2}. Worldwide lightning strokes with 10\,km spatial and 1\,$\mu s$ temporal resolution were provided for the week of September 23, 2019 to match the HL coincident triggers. The required signal travel time for each GLD360 lighting stroke to each detector is computed assuming the minimum frequency-dependent travel speed of $0.75 c$ \cite{KEMP1971919}. This is used as an upper limit for the time difference between signals at each site and the detected lighting. A lower limit is set similarly assuming a signal travel speed of $c$. 

Directly trying to identify coincidences between the GLD360 lightnings and HL coincident triggers on an event-by-event basis leads to a tremendous amount of false coincidences. This is caused by the large amount of GLD360 lightnings as well as HL coincident triggers, in combination with the large time window for identification due to LEMI sampling (4096 Hz) and Omicron time clustering.

To decrease the number of false positives, lightning strokes are broken up into 24 hour segments and clustered by location via K-means clustering\cite{Hartigan1975}. A cluster can roughly be compared to all the lightnings in a thunderstorm. For each cluster of lightnings, an event-by-event comparison is made between the HL coincident triggers and GLD360 strokes. In this comparison a time slide up to $\pm$ 2 s is applied to the GLD360 strokes and the number of coincidences between the GLD360 strokes and the HL coincident lightings is counted as a function of the time slide duration.
In case of a physical coincidence one would expect to observe a large number of coincidences when the time offset is zero and a lower number of coincidences for a non-zero time offset. In the case of purely chance coincidences, no larger amount of coincidences is expected for a zero time offset. By selecting regional lightning clusters where the zero time-offset peak is more than two times the number of background coincidences, we reject coincidences which are likely to be spurious.
Afterwards, the individual coincident lightnings for accepted clusters are selected and these LHO-LLO-Vaisala coincidences, will be referred to as ``lightning-coincident triggers.''

From the two-site coincidence check, we find 314928 total HL coincident triggers. When compared to the 45007030 total GLD360 lightning strokes, we find 268971 total lightning-coincident triggers. We estimate a false-coincident rate of approximately 1 in 1200. This value is estimated from the average rate of background coincidences over the week, measured from the time slides, multiplied by coincidence time window.

In some clusters the background coincidences is about 0.5\%-1\% of the number of lightnings in that given cluster. With the total number of lightnings this would lead to a number of coincidences which is of the same order of magnitude as the amount of HL coincident triggers. This highlights the need of the clustering and time-sliding analysis as a method to reject false coincidences when comparing the GLD360 strokes and the HL coincident triggers.

\begin{figure*}
\centering
\includegraphics[width=0.8\textwidth]{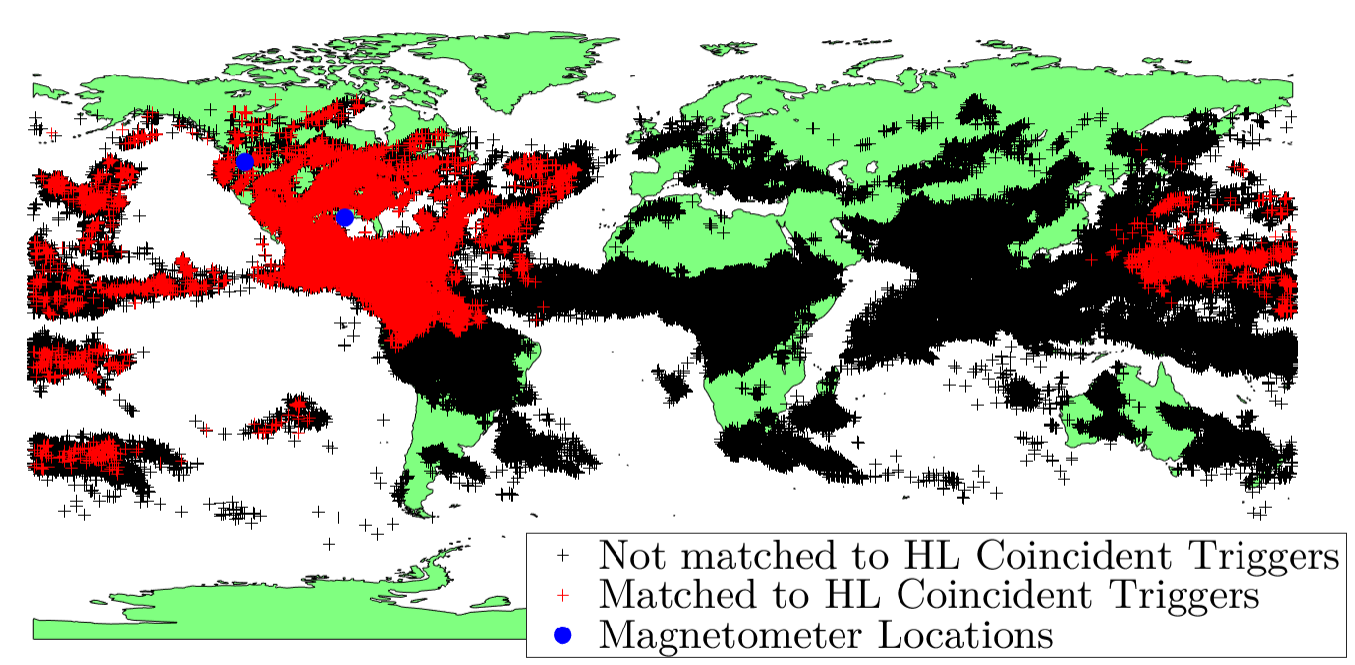}
\caption{Map of all GLD360 lightning strokes during the week of Sept 23, 2019. Red (black) marks are events that did (not) align with HL coincident triggers. The presence of coincident events highlights North America as the primary sensitivity region.}
\label{fig:HL_map}
\end{figure*}

As seen in Fig.~\ref{fig:HL_map}, the coincident lightning strokes were primarily confined to the Americas and Pacific. Over $97\%$ of these were in North and Central America, suggesting that this is the approximate sensitive region for individual lightning signals observed at the LHO and LLO sites.

\begin{figure}
\centering
\includegraphics[width=\linewidth]{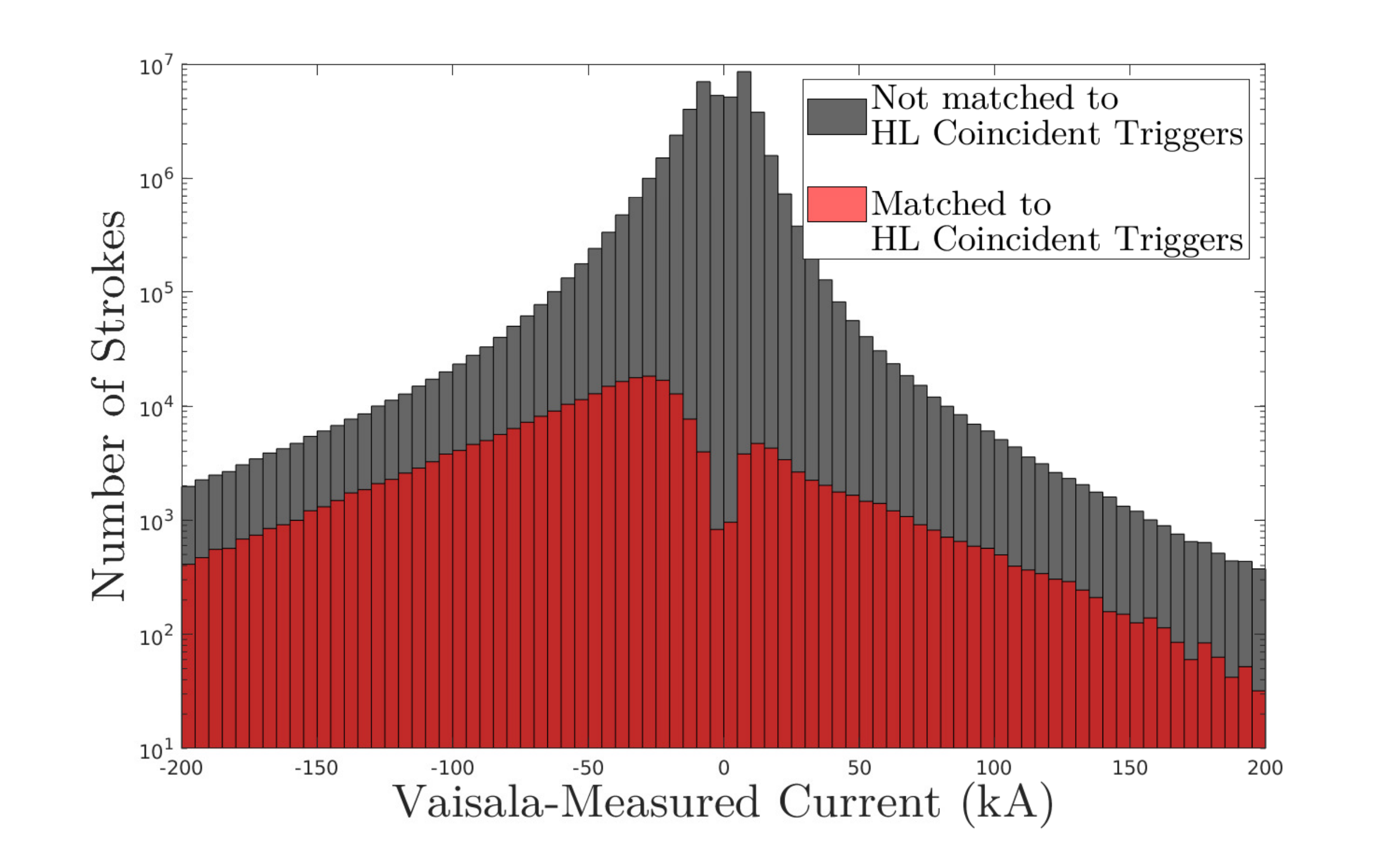}
\caption{Distribution of global lightning stroke currents reported in GLD360. Grey distribution shows the current distribution of all lighting strokes in GLD360 not matched to an HL coincident trigger (corresponding to black locations on Fig \ref{fig:HL_map}). Red distribution shows the current distribution of lightning strokes that were coincident with HL coincident triggers (corresponding to red locations on Fig \ref{fig:HL_map}). The dip at 0 in the matched strokes is likely due to lower current strokes not generating strong enough fields to be witnessed by magnetometers at both LIGO sites.}
\label{fig:lightning_current}
\end{figure}

\begin{figure}
\centering
\includegraphics[width=\linewidth]{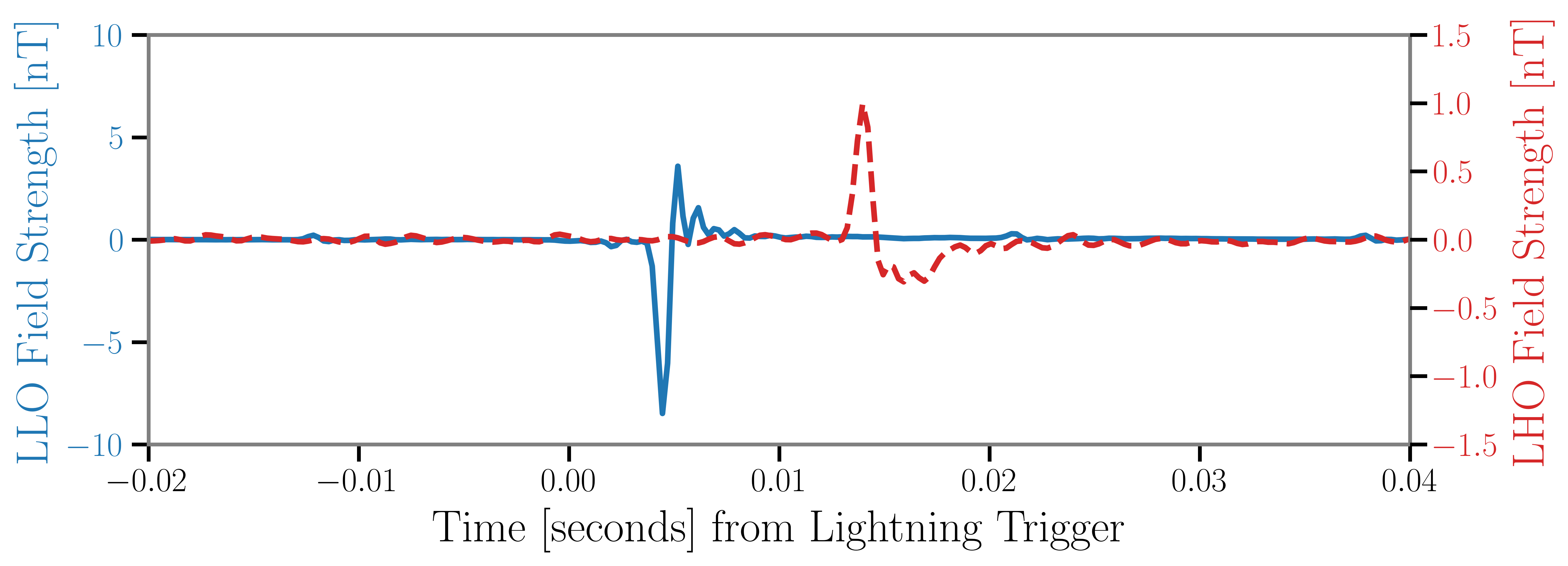}
\caption{Timeseries of a lightning stroke witnessed by magnetometers oriented along the X-arms of each LIGO site. The blue solid line is LLO, and red dashed line is LHO. This is a lightning-coincident trigger linked to a lightning stroke in the Caribbean Sea. 60\,Hz mains lines were removed here.}
\label{fig:magnetic_field_amplitudes}
\end{figure}

\begin{figure}
\centering
\includegraphics[width=\linewidth]{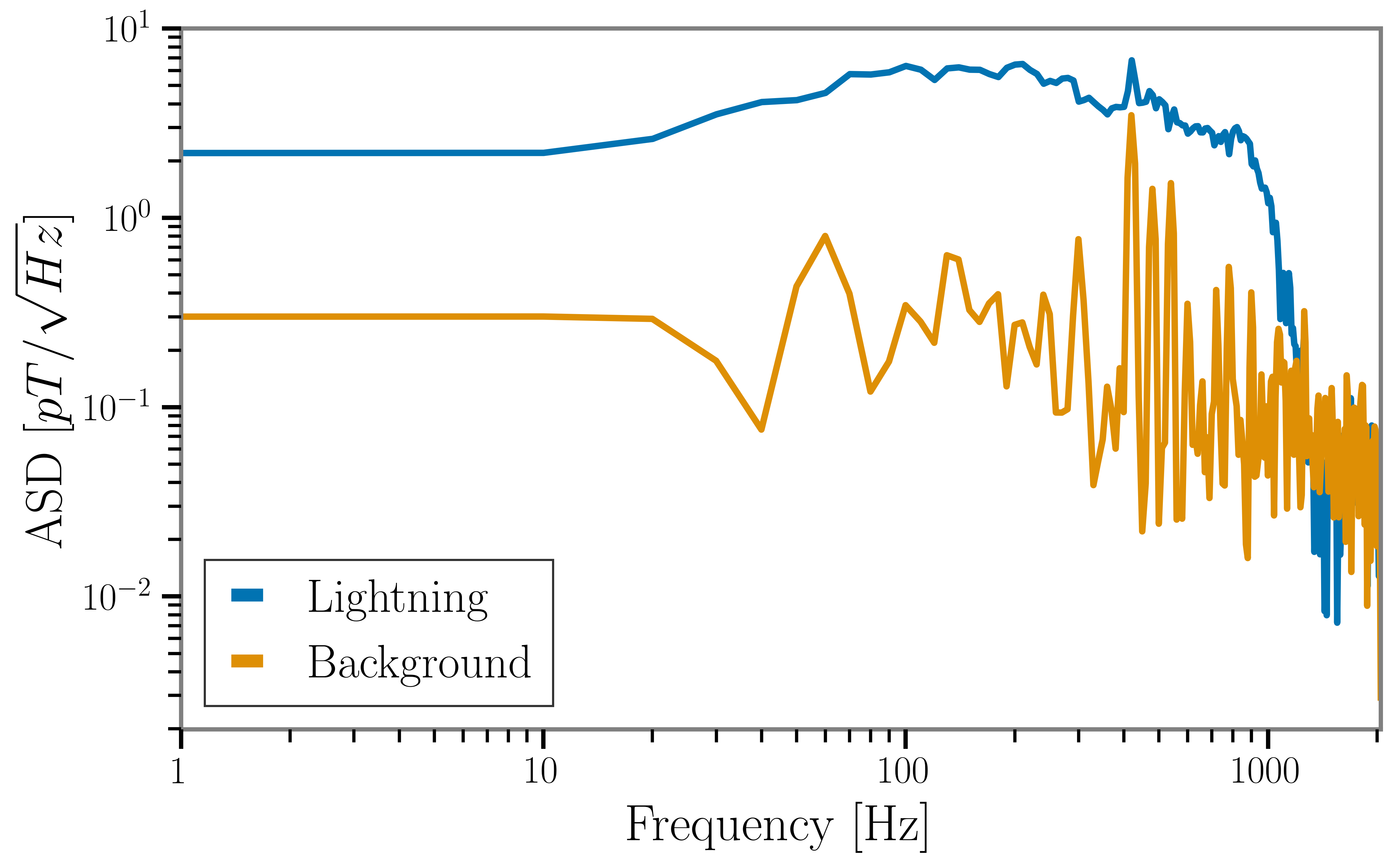}
\caption{Amplitude spectral density from a magnetometer oriented along the LHO X-arm for the event shown in Fig.~\ref{fig:magnetic_field_amplitudes}. Excess energy reaches from below 1 Hz to over 1\,kHz. The blue curve shows 0.1s surrounding the lightning event; the orange curve shows 0.1s of background before the event. These have been corrected for the LEMI response curve, which has decreasing sensitivity at higher frequencies \cite{LEMI_spec_sheet}.}
\label{fig:lightning_spec}
\end{figure}

The distribution of global lightning strokes is generally balanced between positive and negative currents; however, the strokes that match the HL coincident triggers tend heavily toward negative currents (see Fig.~\ref{fig:lightning_current}). Negative currents in the GLD360 are assigned to electric fields pointing toward the Earth~\cite{Said2013}. Strokes matched to HL coincident triggers have a noticeable dip near zero likely due to lower current strokes generating weaker magnetic fields and not being witnessed by magnetometers at both sites. The magnetic field fluctuations from lightning have durations below 10\,ms and amplitudes of a few nT. Fig.~\ref{fig:magnetic_field_amplitudes} shows an example of a signal measured at both LHO and LLO believed to originate from the Caribbean Sea. Spectra of the LHO magnetometers are shown in Fig.\ref{fig:lightning_spec}, where excess amplitude can be seen in the range 1--1000\,Hz.

\begin{figure}
\centering
\includegraphics[width=\linewidth]{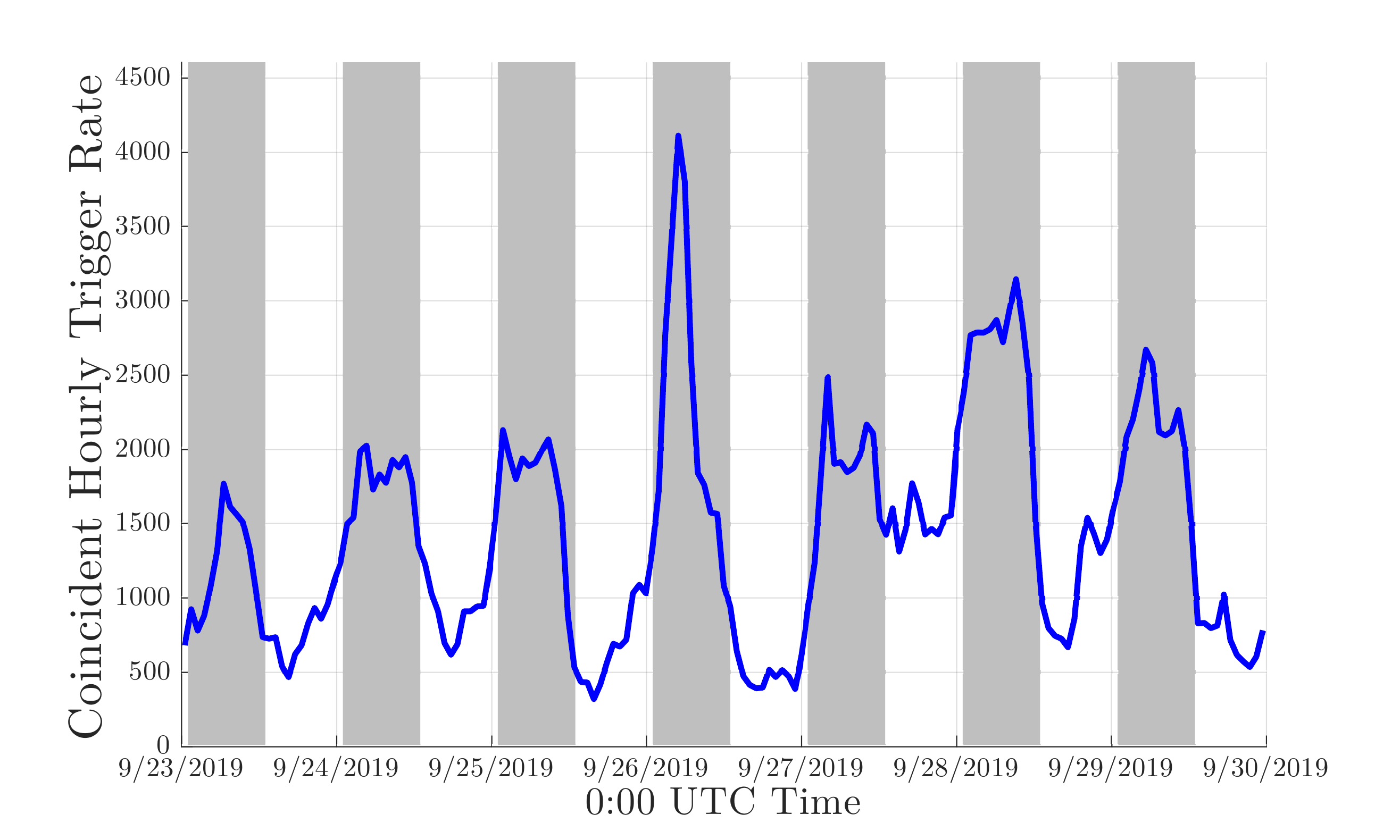}
\caption{Hourly rate of HL magnetic triggers coincident with GLD360 lightning strokes for the week of Sept 23, 2019. Shaded regions mark approximate local night (1-13 UTC) halfway between LHO and LLO. X-axis ticks correspond to 24 hour increments at 0 UTC.}
\label{fig:HL_diurnal}
\end{figure}

\begin{figure}
\centering
\includegraphics[width=\linewidth]{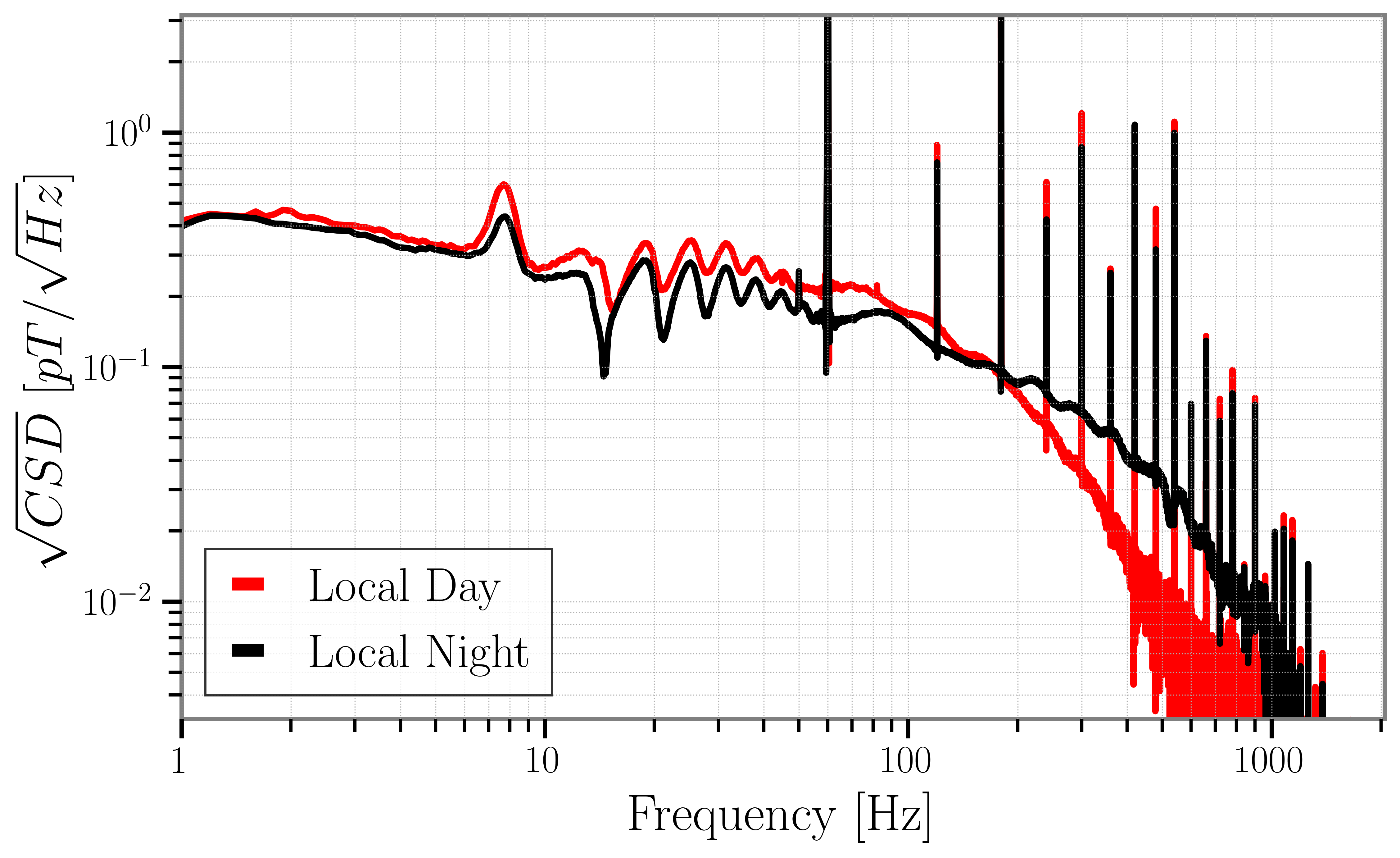}
\caption{Magnetic CSD between the LIGO sites divided into times aligned with day and night in Fig.~\ref{fig:HL_diurnal} for the entire O3 run. At frequencies above $\sim$ 200 Hz the CSD observed during local day (1-13 UTC, red) is smaller than the CSD observed during local night (14-0 UTC, black). The broad peaks below 60Hz are associated to the Schumann resonances. Peaks at multiples of 60\,Hz are due to mains power lines and their harmonics.}
\label{fig:HL_day/night}
\end{figure}

Lightning-coincident triggers are highly diurnal\footnote{Throughout this paper we will use the term diurnal to refer to the daily variations in lightning strikes as is done in literature \cite{blakeslee2014seasonal}.} in occurrence, as shown in Fig.~\ref{fig:HL_diurnal}. The shaded region marks local night in the central United States. The hourly lightning-coincident trigger rate exceeds 2000 coincidences per hour at times. This diurnal rate is most likely a mixture of the local lightning rate (since most strokes are localized spatially and occur during local night) and a diurnal variability in the ionosphere conductivity due to solar radiation~\cite{Simoes2012}. This high diurnal variability in coincident signals suggests that magnetic coherence between sites may also be diurnal. Fig.~\ref{fig:HL_day/night} shows the magnetic cross spectral density (CSD) between the X-arm aligned magnetometers at the LIGO sites when broken up along the day/night cycle shown in Fig.~\ref{fig:HL_diurnal}. Below 200\,Hz, the CSD observed during the day (black line) averages about $20\%$ larger than at night, whereas above 200\,Hz the nighttime CSD (red line) averages about $180\%$ larger than day time.

\subsection{Reducing inter-site magnetic coherence by excluding times with lightning strokes }
\label{sec:lightningCoherence}

\begin{figure}
\centering
\includegraphics[width=\linewidth]{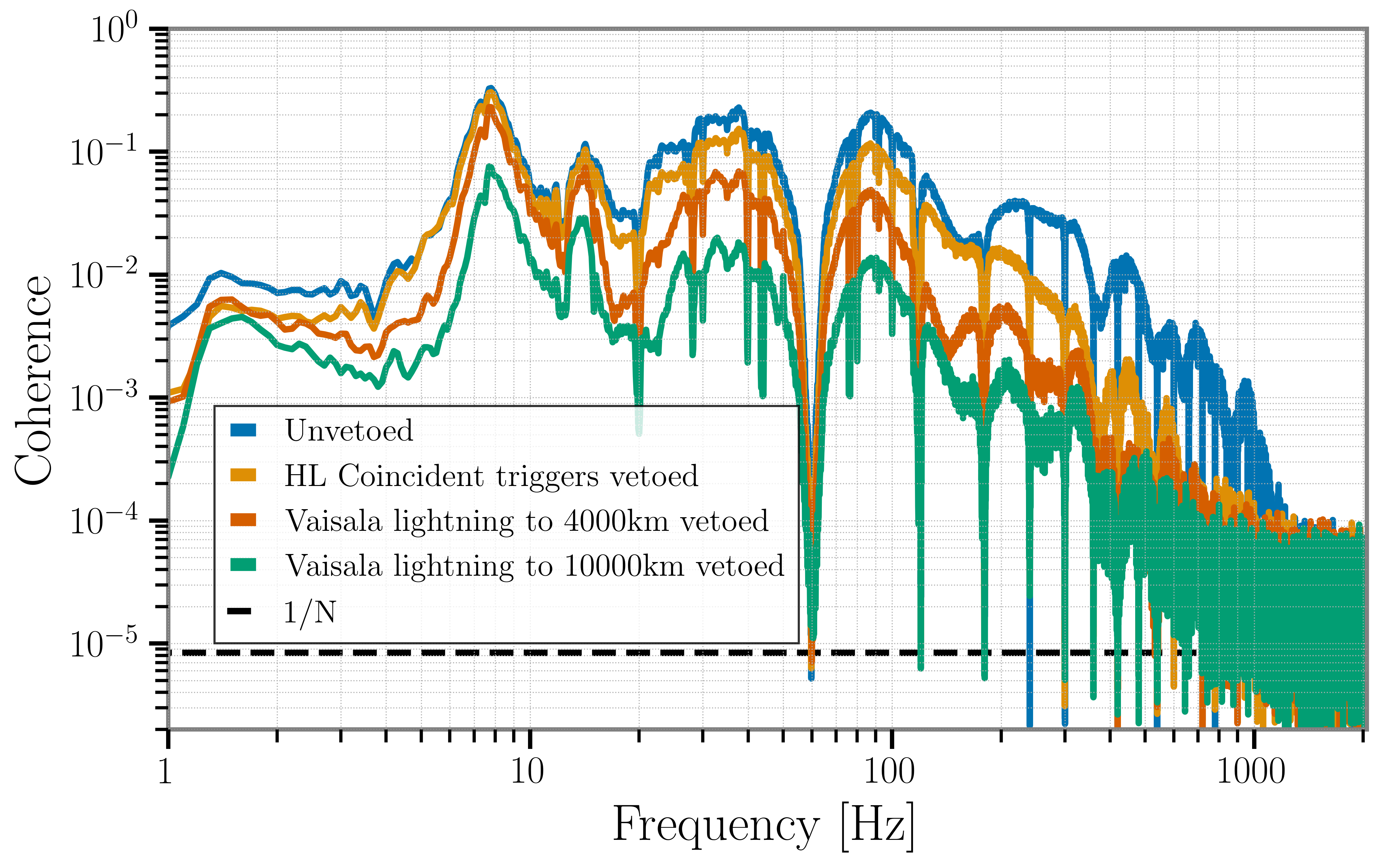}
\includegraphics[width=\linewidth]{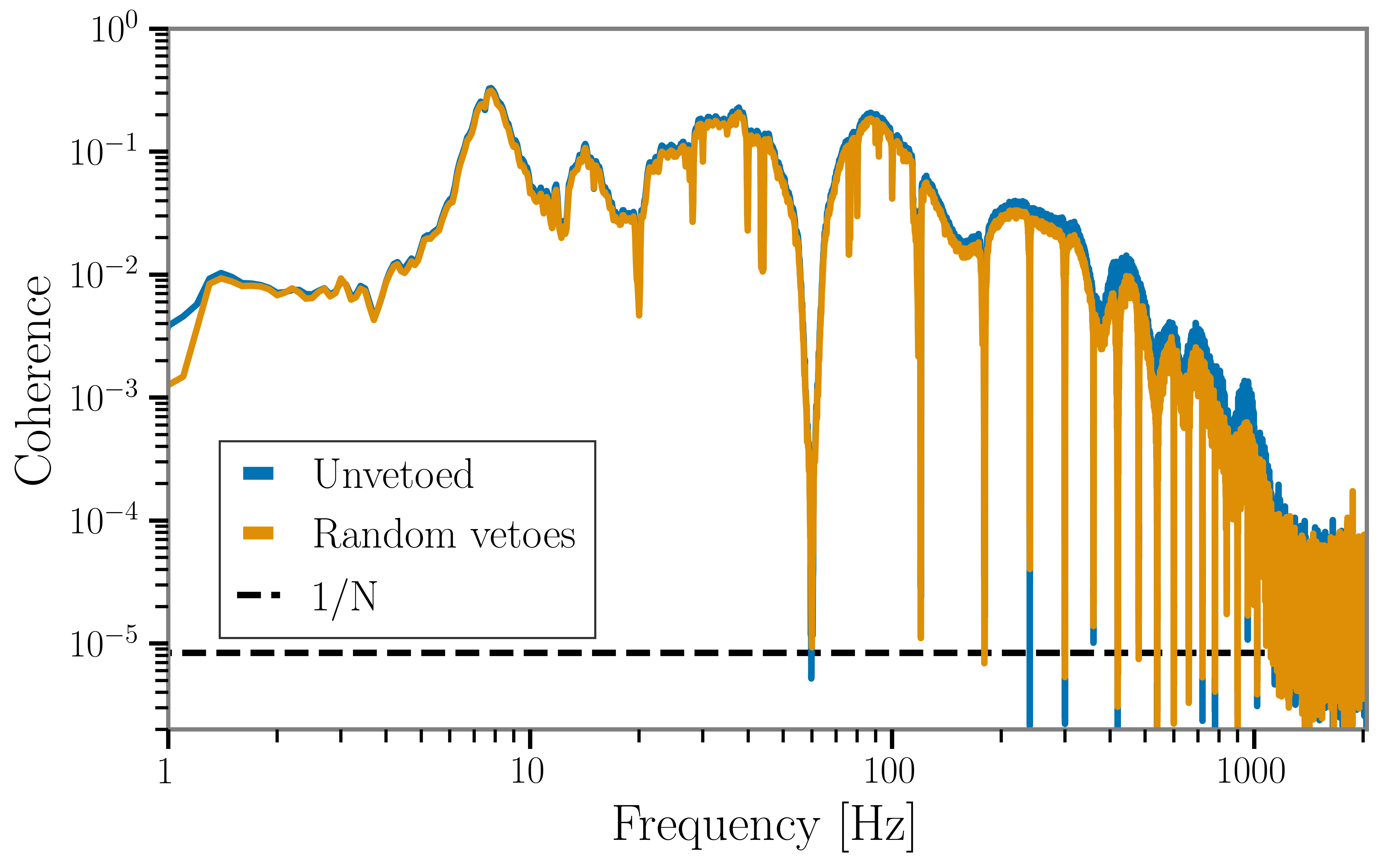}
\caption{Top: Coherence (LHO and LLO magnetometers) spectra for the week of Sept 23, 2019 with varying amounts of lightning events vetoed. In the absence of correlated noise, one expects to be consistent with Gaussian noise given by the number of averages (N) used to make the coherence spectrum, indicated by the black dashed line. Bottom: Randomly applying the same number of trigger vetoes to the data has minimal impact on the coherence spectra suggesting that lightning is the direct source of high frequency coherence. We note that near 1 Hz the coherence with of the data with randomly shifted vetoes is lower. The origin of this is unknown.}
\label{fig:vetoed_cross_correlation}
\end{figure}

We can demonstrate the effect of lightning strokes on inter-site magnetic coherence by vetoing short  time periods that contain individual strokes. Fig.~\ref{fig:vetoed_cross_correlation} shows the coherence between LHO and LLO magnetometers over the course of the week of Sep. 23 2019 with and without vetoes of individual lightning signals. HL coincident trigger times were zeroed in the time-domain with an inverse-Tukey window in one channel (to prevent elevated low frequency coherence due to coincident vetoes) and replaced with Gaussian noise for the duration of the trigger, up to a maximum of 1\,s, modulated by an equivalent Tukey window. This removes a small number of very long outliers with durations up to 1 second. However, the majority of the triggers have durations less than 0.1\,s, with $90\%$ shorter than 0.4\,s. While this reduces the magnetic correlation between LHO and LLO, it did not fully account for the excess, so we also removed times directly tied to GLD360 lighting, accounting for the light travel time from the specific location of the stroke. This was done for GLD360 strokes in a 4000\,km radius circle that best encompassed the sensitivity region shown in Fig.~\ref{fig:HL_map}. We also repeated this, but with all strokes in a 10000\,km radius circle, centered on the same point (far enough to reach to Europe). All GLD360-informed vetoes had durations less than 0.06s. To show that removing magnetic events is the cause of the reduced coherence, we also applied the same number of triggers randomly to the data and recalculated the coherence. The results are shown in the bottom plot of Fig.~\ref{fig:vetoed_cross_correlation}, and we see little to no reduction in the coherence.

The top plot in Fig.~\ref{fig:vetoed_cross_correlation} shows the reduction in coherence as a function of the lightning vetoes used. HL coincident triggers accounted for $0.7\%$ of the globally detected lightning during this time, and represent removing about $7.5\%$ of the observation time. GLD360 strokes in the 4000 km radius circle accounted for $21.8\%$ of globally detected lightning, and the vetoes removed $37.4\%$ of the observation time. The GLD360 strokes in the 10000 km radius circle accounted for $44.9\%$ of global lightning, and the vetoes removed $70.9\%$ of the observation time. Massively increasing the amount of lightning vetoed does reduce the coherence by over an order of magnitude but also removes the majority of the observation time, which is unfeasible in practical application. Whereas going from using HL coincident triggers to using GLD360 detections has a significant impact up to around 300\,Hz -- nearly double the reduction in some frequency ranges -- it has negligible impact on reducing the coherence above a few hundred Hz and requires removing over 9 times as much observation time.


\section{Comparison of  magnetic correlations for different detector locations}
\label{sec:StochPoV}

As part of data quality checks for the search for an isotropic GWB, during O3, the correlations of the magnetic field fluctuations were studied between the LIGO and Virgo observatories. The impact of these magnetic correlations during O3 was reported up to 100Hz \cite{O3Isotropic}; however, this was not done for frequencies above 100\,Hz.

\begin{figure}
\centering
\includegraphics[width=\linewidth]{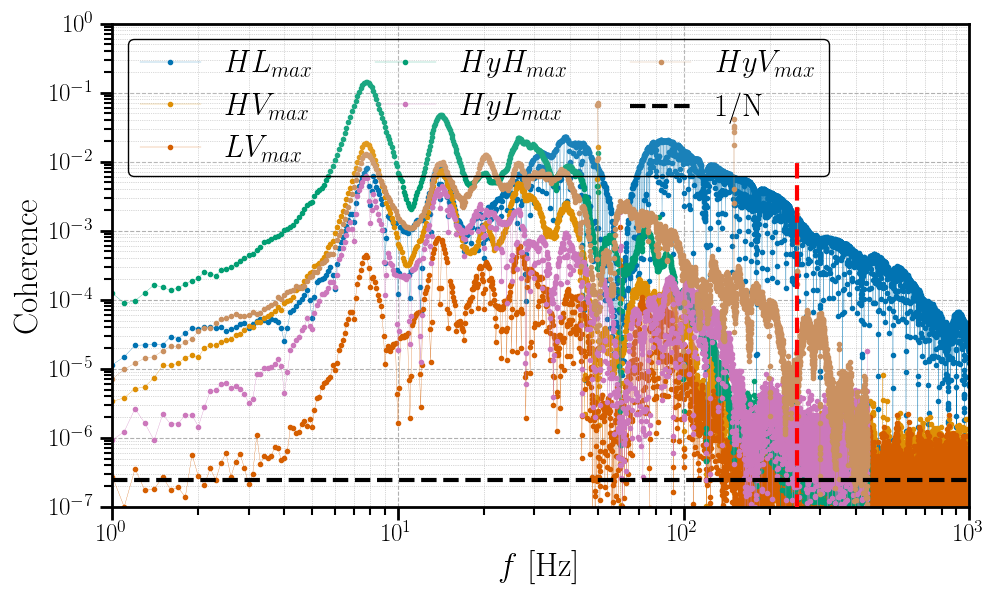}
\caption{Coherence spectrum between HL (blue), HV (orange), LV (red), Hylaty-LHO (green), Hylaty-LLO (purple) and Hylaty-Virgo (brown). At each frequency the maximal coherence is plotted of the four magnetometer pairs for a given baseline. 
In the absence of correlated noise, one expects to be consistent with Gaussian noise given by the number of averages (N) used to make the coherence spectrum, indicated by the black dashed line.
Data spans April 27 2019 00:00:00 UTC to March 27 2019 00:00:00 UTC, consistent with the overlap between LIGO's and Virgo's third observing run (also including data from the commissioning break during October 2019) and the availability of the Hylaty data. Above 250 Hz (red dashed line), the magnetometers at the Hylaty experience significant loss in sensitivity.
}
\label{fig:Hylaty_coh}
\end{figure}

In Fig.~\ref{fig:Hylaty_coh}, the coherences of the magnetic field fluctuations between LHO-LLO, LHO-Virgo and LLO-Virgo are shown (the Hylaty data will be introduced later in this section). The spectra are calculated using data taken between Apr 2 2019 00:00:00 UTC and Mar 27 2020 00:00:00 UTC, consistent with O3 (but also including the data during the commissioning break during October 2019). Furthermore the coherence represented in Fig.~\ref{fig:Hylaty_coh} is, at each frequency, the maximal value from the four possible coherence pairs using the two orthogonal magnetometers at each site (along the x- and y-arms for LHO and LLO and along geographical North-South and East-West for Virgo). 

Some of the Schumann resonances are more coherent in one pair or another. Typically the difference between the minimal and the maximal observed coherence at the different magnetometer pairs for a given baseline is about one order of magnitude or less at a given frequency. In some limited number of cases this is rather two orders of magnitude.

There is a non-zero magnetic coherence above 50\,Hz, where one could expect the strength of the Schumann resonances to be small.
This higher-frequency coherence had not previously been reported in the context of gravitational wave detectors. The observation of this high-frequency inter-site correlation was facilitated by the acquisition, for O3, of  magnetometers placed far from the magnetically noisy buildings.

It is interesting to notice earlier investigations of magnetic correlations between LHO and LLO during the initial detector era reported some excess coherence between 50--100\,Hz \cite{Thrane:2013npa}. This magnetic coherence during the fifth science run (S5) is rather minimal above 50\,Hz and does not extend above 100\,Hz. Since the integration time used in the S5 analysis, is quite similar to the time used in Fig.~\ref{fig:Hylaty_coh}, coinciding with O3, the difference is entirely due to the dedicated magnetometers used during O3. During S5 the magnetometers were measuring inside the buildings and are therefore measuring in a magnetically noisier environment, leading to the magnetometers mainly being dominated by local magnetic sources.

\begin{figure}
\centering
\includegraphics[width=\linewidth]{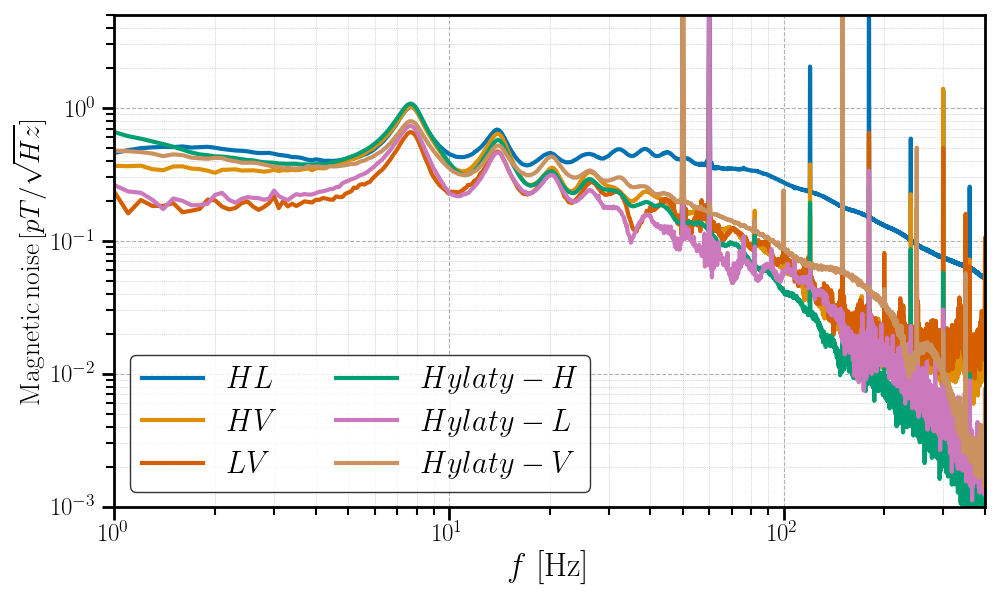}
\caption{CSD$^{1/2}$ (as given by Eq. \ref{eq:OmniDirecCSD}) of HL, HV, LV, Hylaty-LHO, Hylaty-LLO and Hylaty-Virgo. Data is used from April 27 2019 00:00:00 UTC to March 27 2019 00:00:00 UTC, consistent with the overlap between LIGO's and Virgo's third observing run (also including the data from the commissioning break during October 2019) and the availability of the Hylaty data. Above 250 Hz (red dashed line) the magnetometers at the Hylaty experience significant loss in sensitivity.
The broad peaks below 60Hz are associated to the Schumann resonances.}
\label{fig:Hylaty_stochCSD}
\end{figure}

Fig.~\ref{fig:Hylaty_stochCSD} reports the CSD between the different magnetometers for the HL, HV and LV baselines -- that is a pair of the GW detectors -- of data taken between Apr 2 2019 00:00:00 UTC and Mar 27 2020 00:00:00 UTC. Here, we show the quadratic sum of the four different individual CSD combinations, as represented in Eq. \ref{eq:OmniDirecCSD}. Furthermore we use the modulus of the individual magnetic CSD instead of only the real part. This choice is made since it is the most conservative option we can choose:
\begin{equation}
\label{eq:OmniDirecCSD}
\begin{aligned}
                CSD_{IJ} =& \frac{2}{T} \left[ \right.  |\Tilde{m}^*_{I_1}(f)\Tilde{m}_{J_1}(f)| ^2 + |\Tilde{m}^*_{I_1}(f)\Tilde{m}_{J_2}(f)|^2  \\
             &+ |\Tilde{m}^*_{I_2}(f)\Tilde{m}_{J_1}(f)| ^2 +  |\Tilde{m}^*_{I_2}(f)\Tilde{m}_{J_2}(f)|^2 \left.\right]^{1/2} ,        
\end{aligned}
\end{equation}
with $\Tilde{m}^*_{I_1}(f)$ the Fourier transform of the the magnetic data $m_{I_1}(t)$ measured by sensor 1 located at GW detector $I$.
The factor of $2/T$ is a normalisation constant of the Fourier transform, where $T$ is the time duration of the data segment when applying the Fourier transform. In the remainder of the paper we will always use this conservative estimate of the CSD when discussing the magnetic CSD.

Both the magnetic coherence (Fig.~\ref{fig:Hylaty_coh}) and the CSD (Fig.~\ref{fig:Hylaty_stochCSD}), fall off more rapidly with increasing frequency in the case of HV and LV compared to HL. The fast decline in magnetic CSD for the HV and LV baselines can be understood from the increased attenuation of magnetic field fluctuations caused by lightnings for larger distances. Namely the distance between LHO and LLO is about 3000 km, whereas the distances between Virgo and LHO, LLO are respectively $\sim$ 8800 km and $\sim$ 8500 km. Secondly this is also in line with the elevated thunderstorm activity in the Americas compared to Europe. The rate of lightning near LLO tends to be greater than the rate in Europe, and the rate around LHO is lower than around Virgo and comparable to Poland \cite{vaisala_lightning_report}.

To further test this hypothesis, we will use data of another magnetometer station. The station we use is part of the WERA project \cite{WERAProj}: the Hylaty station located in the Bieszczady Mountains in Poland \cite{PolishMag}. 
The distance between the Hylaty station LHO, LLO and Virgo is respectively $\sim$ 8800 km, $\sim$ 9000 km and $\sim$ 1100 km.
The station is equipped with two magnetometers, oriented along North-South and East-West and is located in a very quiet location to measure ELF magnetic fields. The magnetometer response, as a function of frequency, is quite flat up to 250\,Hz, but afterwards starts to decline rapidly with only 50\% response at 262 Hz. Therefore we will focus in the following analysis on the frequency region 1--250\,Hz.

The coherence and CSD between the magnetometers located at Hylaty and GW detectors are shown in respectively Fig. \ref{fig:Hylaty_coh} and Fig. \ref{fig:Hylaty_stochCSD}. 
First of all, we notice the coherences between different baselines span several orders of magnitude, even at the Schumann frequencies. Above 100\,Hz, the HL-baseline has the highest coherence followed by the Hylaty-Virgo baseline. HV and LV have the lowest coherence of all.
When looking to the CSD, we notice the amplitude of the CSD at the fundamental Schumann mode (7.8\,Hz) ranges from 0.4 pT$^2$/Hz to 1 pT$^2$/Hz. Above 100Hz the CSD of HL is the largest by one order of magnitude. The baseline CSDs in descending order are (above 100 Hz): $\mathrm{CSD}_{\mathrm{HL}} > \mathrm{CSD}_{\mathrm{Hylaty-V}} > \mathrm{CSD}_{\mathrm{HV}} \approx \mathrm{CSD}_{\mathrm{LV}} > \mathrm{CSD}_{\mathrm{Hylaty-H}} \approx \mathrm{CSD}_{\mathrm{Hylaty-L}}  $.

This can be explained by the characteristics of the propagation of electromagnetic waves in the Earth's atmosphere and the location of the main thunderstorm regions. 

The magnetometers located at LHO-LLO and Hylaty-Virgo have the highest CSDs. The distance between the magnetometers for other baselines is 2.5 to 9 times larger, leading larger attenuation of the electromagnetic waves. The significantly higher coherence and CSD in the HL baseline compared to the Hylaty-Virgo baseline can be explained by the main thunderstorm regions. Whereas LHO and LLO are located in a (very) active thunderstorm region, Hylaty and Virgo are further away from the main thunderstorm regions, located in America, Africa and Asia.

The observed behaviour discussed in this section is consistent with what is expected from our hypothesis postulated in Sec. \ref{sec:IndivLightnings}, i.e. the superposition of individual lightnings is the likely source to explain the observed coherence and cross-correlation between the magnetometers at the GW detectors during O3.


\section{Impact on gravitational wave searches}
\label{sec:ImpactfutureSearches}

In this section we will study the impact of the (superposition of) individual lightning strokes on searches for GWs. The projection of magnetic noise onto the sensitivity of a GW analysis is typically called a magnetic budget. By constructing this magnetic budget we make predictions at which level these magnetic disturbances couple to our interferometric detectors. In Sec. \ref{sec:magneticcoupling} we will discuss the mechanisms of how magnetic fields can couple to the GW detectors. In Sec. \ref{sec:impactGWB} we will discuss the impact on the search for anisotropic GWB, whereas Sec. \ref{sec:impactBurst} focuses on the impact of the individual lightning strokes on transient GW searches. 

\subsection{Coupling of magnetic fields to GW interferometric detectors}
\label{sec:magneticcoupling}

Magnetic fields can couple to a GW interferometric detector by acting on the electromagnetic actuators located on the test masses and/or their suspensions \cite{Nguyen_2021,galaxies8040082,Cirone_2018,Cirone_2019}. This is typically the dominant effect at low frequencies. At higher frequencies other mechanisms (sometimes also specific to a certain interferometer) induce magnetic coupling to the interferometer, such as interaction with signal cables or by acting on optical components such as suspended benches \cite{Cirone_2019,galaxies8040082}. 

To measure the coupling of magnetic fields to the interferometer, magnetic fields are generated in the central buildings on a weekly basis (during observing runs) \cite{Nguyen_2021,galaxies8040082}. 
These are measured by injecting magnetic fields ,typically a range of different frequency sine-wave signals, which are observed by witness magnetometers. However these magnetometers are not necessary positioned in the exact locations where the magnetic fields couple to the interferometer, possibly leading to a difference between the measured and true magnetic coupling.
By creating magnetic fields many times larger than the typical ambient fields, one can study the level of magnetic coupling to the interferometer.
However these weekly injections do not target coupling sensitive locations, but rather the majority of the building. Other locations can couple more significantly to the interferometer. To this extent at the start of the run\footnote{Given the emergency shutdown due to the Covid-19 pandemic only limited injection tests were performed at the end of O3. Since these were consistent with the sitewide coupling measured at the start of the run, no further measurements were made.} a `sitewide' injection was performed for LHO and LLO at many possible coupling locations, as is described in \cite{Nguyen_2021}\footnote{Please note that in \cite{Nguyen_2021} they use `composite' coupling instead of sitewide coupling to refer to this measurement.}. The resulting sitewide coupling consist out of the highest value in every frequency bin from all the different physical locations at which fields were injected. In this paper we use the sitewide coupling measured at the start of O3, rather than the weekly measurements, to have a conservative estimate.

At frequencies above several hundreds of Hz, the effect from the injected fields is often not strong enough to be observed in the GW sensitive strain channel. In that case an upper limit of the magnetic coupling function is provided.
At Virgo it was possible to measure the high frequency coupling during some weeks \cite{galaxies8040082}, while at the LIGO detectors, only upper limits were placed above $\sim$ 200Hz-300Hz.

The coupling function discussed above describes the level of magnetic coupling in the differential arm length (DARM) \footnote{DARM is the GW sensitive channel of our interferometer.}, based on the magnetic field inside the experimental building(s). Therefore we will use the term inside-to-DARM magnetic coupling function.

The magnetometers we use to measure inter-site coherence are located outside the experimental buildings to measure the weak ambient fields. Therefore we also need to take into account the effect of the building. The reduction/amplification describing the projection from magnetic fields outside-to-inside the experimental building is called the outside-to-inside magnetic coupling function.

Previous estimates of this outside-to-inside magnetic coupling function were calculated based on using a coil generating magnetic fields outside the central building at LHO \cite{O3Isotropic}. However, the fields generated by this coil differ from lightning strokes in time-scale and field uniformity, i.e. magnetic fields from lightning strokes have shorter timescales and larger magnetic field uniformity. Therefore the outside-to-inside magnetic coupling was measured using the magnetic fields generated by distant lightning strokes. This method is explained in more detail in Appendix \ref{appendix:outsideInsideTF}.
A scalar outside-to-inside magnetic coupling function was calculated for magnetic fields from lightning for LLO. The distribution was found to be log-normal with a mean of $0.7^{+0.4}_{-0.3}$ nT/nT. Future studies should aim to measure the outside-to-inside magnetic coupling function for lightning fields at the different sites.

\subsection{Impact on the search for an isotropic GWB}
\label{sec:impactGWB}

The isotropic GWB analysis tries to measure $\Omega_{GW}$\cite{PhysRevD.46.5250,PhysRevD.59.102001,LivingRevRelativ20}, the energy density $\text{d}\rho_{GW}$ contained in a logarithmic frequency interval $\text{d}lnf$, divided by the critical energy density $\rho_c$ needed for a flat universe:
\begin{equation}
    \label{eq:omegaGW}
    \Omega_{GW} = \frac{1}{\rho_c} \frac{\text{d}\rho_{GW}}{\text{d}lnf}
\end{equation}
In the absence of correlated noise, the cross-correlation statistic $\hat{C}_{IJ}(f)$ (Eq. \ref{eq:Cij}) for detectors $I$ and $J$, is an unbiased estimator of $\Omega_{GW}$ \cite{PhysRevD.59.102001,LivingRevRelativ20}
\begin{equation}
    \label{eq:Cij}
    \hat{C}_{IJ}(f) = \frac{2}{T} \frac{Re[\Tilde{s}^*_I(f)\Tilde{s}_J(f)]}{\gamma_{IJ}(f)S_0(f)}.
\end{equation}
An explicit expression for the contribution of the correlated noise that biases this estimator can be found in Eq. 12 of \cite{PhysRevD.102.102005}. 
$\Tilde{s}_I(f)$ is the Fourier transform of the time domain data $s_I(t)$ measured by interferometer $I$, $\gamma_{IJ}$ the normalized overlap reduction function~\cite{PhysRevD.46.5250} which encodes the baselines geometry and $S_0(f)$ is given by $S_0(f)=(3H_0^2)/(10\pi^2f^3)$, where $H_0$ is the Hubble-Lema\^\i tre constant. $T$ is the total observation time of your data taking period. However, if you are analysing the data in separate segments this becomes the duration of the time segments used.

In the small signal-to-noise ratio limit for the GWB, the uncertainty on $\hat{C}_{IJ}(f)$ is given by Eq. \ref{eq:sigmaGWB}, where $\Delta f$, is the frequency resolution \cite{PhysRevD.59.102001,LivingRevRelativ20}.
\begin{equation}
    \label{eq:sigmaGWB}
    \sigma_{IJ}(f) \approx \sqrt{\frac{1}{2T\Delta f}\frac{P_I(f)P_J(f)}{\gamma_{IJ}^2(f)S_0^2(f)}}
\end{equation}
Equivalent to the cross-correlation statistic (Eq. \ref{eq:Cij}) one can construct a magnetic cross-correlation statistic given by Eq. \ref{eq:C_Mag} \cite{Thrane:2013npa,Thrane:2014yza}.
\begin{equation}
    \label{eq:C_Mag}
        \hat{C}_{mag,IJ}(f) = |\Kappa_I(f)||\Kappa_J(f)|  \frac{2}{T} \frac{Re[\Tilde{m}^*_I(f)\Tilde{m}_J(f)]}{\gamma_{IJ}(f)S_0(f)},
\end{equation}
where $\Kappa_I(f)$ describes the coupling from magnetic fields to interferometer $I$ and $\Tilde{m}_I(f)$ is the Fourier transform of the time domain data $m_I(t)$ measured by a magnetometer at site $I$. $T$ is the duration of the segments used when Fourier transforming the magnetic data.

When analysing data in search for an isotropic GWB, one typically constructs the magnetic cross-correlation statistic $\hat{C}_{mag,IJ}(f)$ to investigate if the observed magnetic fields might result in correlated noise in the analysis. 
The magnetic coupling functions $\Kappa_I(f)$ are the product of the outside-to-inisde and inside-to-DARM magnetic coupling functions.
The latter is measured by injecting magnetic fields with known amplitude and frequency and observing their impact on the output in the GW data $s_I(t)$\cite{galaxies8040082,Nguyen_2021,Cirone_2018,Davis_2021,alog:mag_fac,alog:O3PEMInjections, alog:WeeklyMagneticInjections}.

At all sites $I$, the environmental magnetic field outside the buildings is measured in the local horizontal plane by two magnetometers, perpendicular to each other: $m_{I_1}$ and $m_{I_2}$. To be conservative, we will use the quadratic sum of the four different combinations as well as the modulus of the magnetic cross spectral density -- see Eq. \ref{eq:OmniDirecCSD} -- instead of only the real part as in Eq. \ref{eq:C_Mag}. This leads to Eq. \ref{eq:C_Mag_final}.

\begin{equation}
    \label{eq:C_Mag_final}
    \begin{aligned}
        \hat{C}_{mag,IJ}(f) &= |\Kappa_I(f)||\Kappa_J(f)|  \frac{CSD_{IJ}}{\gamma_{IJ}(f)S_0(f)}\\
    \end{aligned}
\end{equation}

The method used to construct the magnetic budget presented here is different to the budget presented by the LIGO, Virgo and KAGRA collaborations in their search for an isotropic GWB~\cite{O3Isotropic}. 
There are four main differences, the first being the larger frequency range from 20\,Hz to 675\,Hz which is the entire frequency range having regular measurements of the inside-to-DARM magnetic coupling at all three sites. A linear interpolation as a function of frequency is performed.
Secondly, the LVK collaborations used the magnetic spectrum obtained from one (the worst) pair e.g. $Re[\Tilde{m}^*_{I_2}(f)\Tilde{m}_{J_1}(f)]$~\cite{O3Isotropic}, whereas here we use the more conservative $CSD_{IJ}$, combining the modulus for all four CSD-pairs for one baseline. In ~\cite{O3Isotropic}, an additional factor of 2 is included because of their choice of one single direction compared to the total CSD.
Thirdly, in \cite{O3Isotropic} they use an approximate outside-to-inside transfer function of 1, based on measurements made using injection coils at LHO. Here we will use the measurement for LLO described in Appendix \ref{appendix:outsideInsideTF}, based on lightning-generated magnetic fields instead of coil-generated fields, with a resulting value of $0.7^{+0.4}_{-0.3}$ nT/nT. For the budget we will assume both LHO and Virgo have the same outside-to-inside coupling function as LLO. 

In the future dedicated studies at the other sites should give accurate site-dependent measurements. 
Fourth and finally, they used the weekly magnetic coupling measurements at each site and computed a budget for every week of O3. Here we will rather use the sitewide coupling functions measured for LHO and LLO since they are more conservative. For Virgo we will use the average of the weekly measurements in the central building, which has the largest coupling. A downside of the site-wide coupling with respect to the weekly measurements is that for LHO the frequency resolution is worse compared to the resolution of the weekly injections.

Similar to the budget presented in \cite{O3Isotropic} the three separate magnetic budgets for each baseline are combined weighing them by the sensitivity of the respective baseline to a GWB:
\begin{equation}
\label{eq:totMagBudget_weights}
\begin{aligned}
        \hat{C}_{\rm mag}&=\sum_{IJ} w_{IJ}(f) \hat{C}_{{\rm mag},IJ}(f), \text{ where} \\ w_{IJ}(f)&=(\sigma_{IJ}(f)/\sigma(f))^{-2} \text{ and} \\
        \sigma^{-2} &= \sum_{IJ} \sigma_{IJ}^{-2}.
\end{aligned}
\end{equation} 
This weighing method was first proposed in \cite{PhysRevD.102.102005}. Given that the HL-baseline is the most sensitive pair to GWs, its magnetic budget will dominate the total budget presented here, except around the zeros of it's GW overlap reduction function.

\begin{figure*}
\centering
\includegraphics[width=0.8\linewidth]{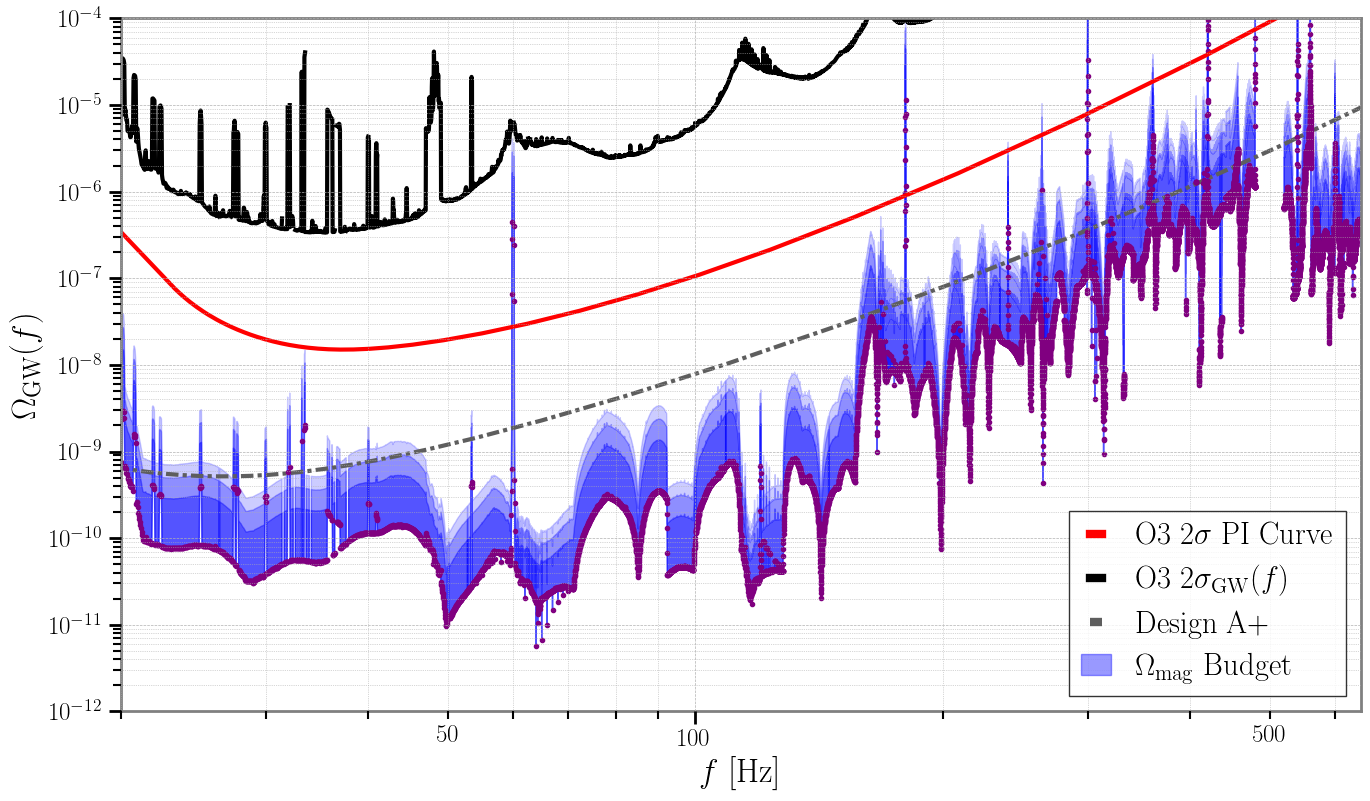}
\caption{Magnetic noise budget represented by the blue band. The purple dots represent the budget without any errors included. The dark and lighter blue bands represent the upper 1$\sigma$-3$\sigma$ uncertainty as described in Appendix \ref{appendix:errorpropagation}. No error is included for the weekly variation, however as explained in the text this effect was found to be minimal.
The lower errors are not shown in this figure since the error propagation as described in Appendix \ref{appendix:errorpropagation} leads to negative lower limits.
Also represented are the O3 sensitivity for narrowband features for an isotropic GWB, given by its standard deviation $\sigma(f)$; the O3 broadband sensitivity, given by its power-law integrated (PI) curve (red); and the broadband sensitivity expected to be reached with the LIGO A+ and Advanced Virgo Plus network, Design A+ (grey dot-dashed).}
\label{fig:GWB_MAG_budget}
\end{figure*}

The darkest blue band in Fig. \ref{fig:GWB_MAG_budget} reports the upper 1$\sigma$ uncertainty band. The lighter blue bands respectively indicate the upper 2$\sigma$ and 3$\sigma$ uncertainty. No lower errors are shown, since they are found to be negative.
In Appendix \ref{appendix:errorpropagation} the error propagation of the budget is discussed in detail. This includes the uncertainty of the outside-to-inside coupling as well as a factor 2 of uncertainty related to inside-to-DARM magnetic coupling function, given the limitations of the inside-to-DARM coupling function measurement \cite{Nguyen_2021}. The variation of the weekly measurements was observed at the central building of each site. However the sitewide coupling functions used for LHO and LLO rely on measurements on different physical locations, which were only performed at the start and end of O3. Given there is no guarantee that these other coupling locations have the same weekly variation, no error for the weekly variation was taken into account. The weekly variation was measured, as described in Appendix \ref{appendix:errorpropagation}, and was found to have a minimal effect on the budget presented in Fig. \ref{fig:GWB_MAG_budget}.
Ideally, the weekly variation of the sitewide coupling would also be measured in the future such that this error can be taken into account.

We want to point out that much of the topology of the budget is due to the limited frequency resolution of the measured coupling function and the linear interpolation in between points. The peaks and dips near 50 Hz and above are a clear example. Ideally future measurements of the inside-to-DARM magnetic coupling function would have finer frequency resolutions. This can addressed by performing broadband noise injections, rather than injecting sine-wave signals, which is recommended for future observing runs.

We conclude that there was no magnetic contamination in the search for an isotropic GWB during O3, which is consistent with the data being in agreement with Gaussian noise. The same conclusion was made based on the magnetic budget by the LVK collaborations in the frequency range 20--100\,Hz \cite{O3Isotropic}.
Also at frequencies between 100--675\,Hz, there was no magnetic contamination, neither narrowband, nor broadband. Concerning narrowband magnetic contamination, one should compare the upper range of the magnetic budget and its narrow features with the standard deviation of the search for an isotropic GWB $\sigma(f)$. We note that the loudest peaks (60 Hz and harmonics) linked to the US power mains are excluded from the analysis \cite{O3Isotropic}.
The blue band of the magnetic budget being below the powerlaw integrated (PI) sensitivity curve (red line) \cite{Thrane:2013oya} implies there is no broadband magnetic contamination.

In case the correlations from the magnetic field fluctuations, both amplitude and frequency behavior, as well as the magnetic coupling (outside-to-inside as well as inside-to-DARM) remains the same for future observing runs, there is a risk magnetic contamination might affect the search for an isotropic GWB when LIGO and Virgo reach respectively the LIGO A+ and Advanced Virgo Plus sensitivities \cite{APlussReference}. These sensitivities are planned to be reached later this decade.
Up to $\sim$50 Hz, the 2$\sigma$ budget contour touches the PI curve indicating there is a non-negligible risk that magnetic noise might bias or affect the search for an isotropic GWB. 
From $\sim$ 160 Hz, there is again a non-negligible possibility of significant magnetic contamination. However, we note that the values of the inside-to-DARM magnetic coupling at these and higher frequencies are often not measured but are rather upper limits.

Based on this budget we believe more work in the upcoming observing run(s) is needed. 
First, we recommend to increase the magnetic injection strength at frequencies above $\sim$160 Hz to either measure the inside-to-DARM magnetic coupling or push down the upper limits by a factor of 2-3 at each interferometer.
Second, one should strive to decrease the variability of the outside-to-inside magnetic coupling with location in the building and the direction to the lightning. In order to reduce the coupling variability, it may be necessary to reduce the lightning-induced currents on the beam tube that pass through the buildings (see Appendix \ref{appendix:outsideInsideTF}). Furthermore ideally the outside-to-inside magnetic coupling would be measured at each site.
Third, some improvement in the roughly factor of two uncertainty in the inside-to-DARM coupling functions may be possible, though this may be difficult because magnetometers can not be placed exactly at the coupling sites \cite{Nguyen_2021}. The budget would also benefit from a finer frequency resolution of the measurements of the inside-to-DARM coupling functions.
Finally, instead of a combined budget for an entire observing run, the time-dependence of the magnetic coupling and its effect on the search for a GWB should be studied in more detail. This ideally would entail performing weekly (nearly) sitewide injections.
Also the diurnal (and seasonal) effects on the amplitude of the magnetic field fluctuations could be used to differentiate between correlated noise and a GWB signal.
Furthermore we want to point out that in the past several efforts have been studied to validate the detection of a GWB in the case of correlated noise. This includes the use of GW-goedesy \cite{GWGeodesy,PhysRevD.105.082001}, joined Bayesian parameter estimation \cite{PhysRevD.102.102005} and Wiener filtering \cite{Thrane:2013npa,Thrane:2014yza,10.1088/0264-9381/33/22/224003,PhysRevD.97.102007}. However these works have mainly focused on the Schumann resonances.

The effect of correlations from magnetic field fluctuations on future generation GW interferometric detectors has been studied for the Einstein telescope \cite{janssens2021impact}. One should also be aware that the high frequency correlated noise can be larger for detectors near large thunderstorm regions and/or mutual closely located detectors.

\subsection{Impact on the transient GW searches}
\label{sec:impactBurst}

\begin{figure*}
\centering
\includegraphics[width=0.8\linewidth]{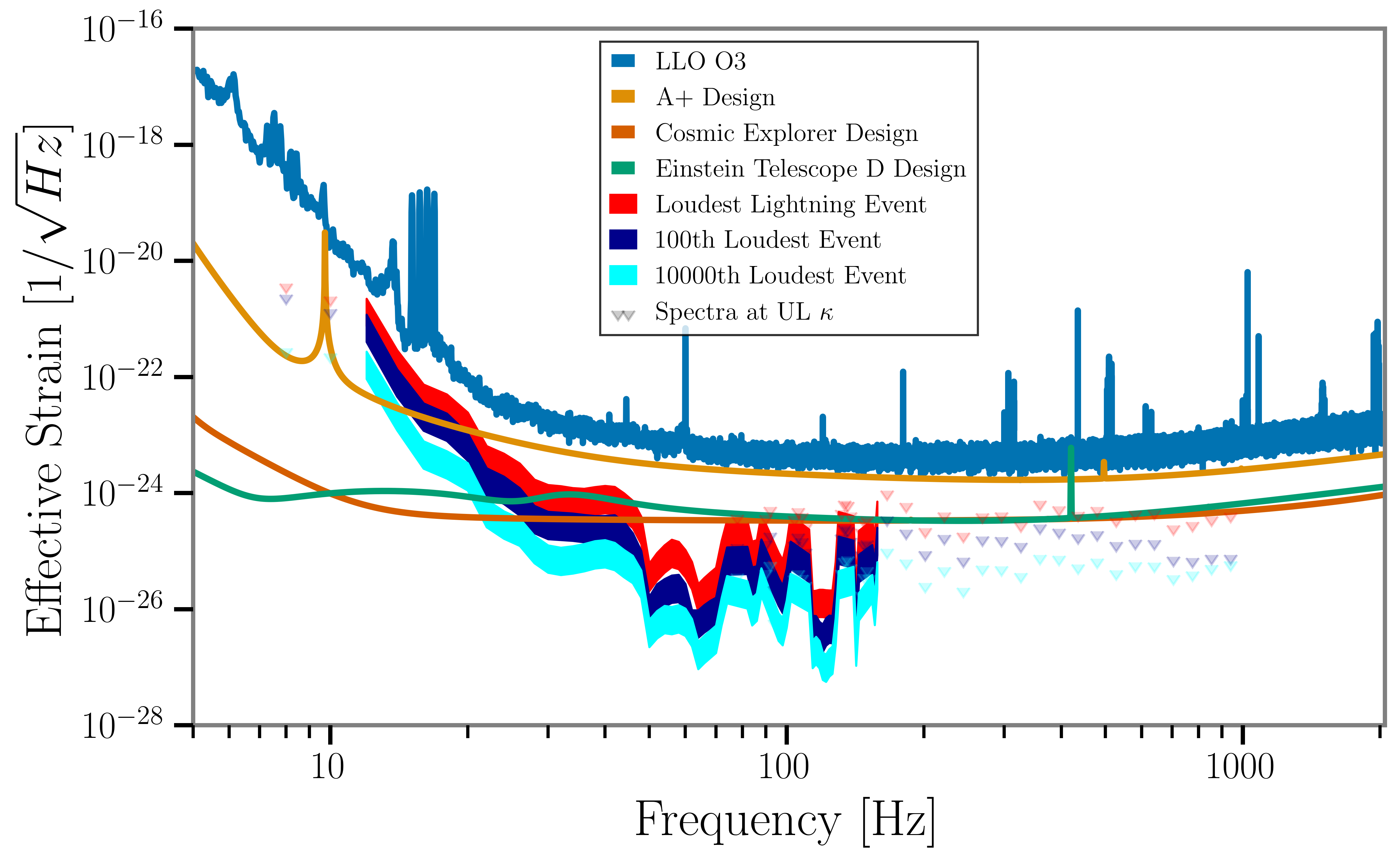}
\caption{Estimates of the strain noise produced by individual lightning strokes for LIGO and, assuming the same magnetic coupling, for future detectors. The wide bands were made by projecting LEMI magnetic field spectra from the 1st, 100th, and 10000th highest amplitude lightning strokes detected by Omicron projected into the main GW strain channel through the outside-to-inside and inside-to-DARM magnetic coupling functions as measured at LLO. Spectra represent 0.5s around lightning event. Shaded regions represent the 68\% confidence interval for the spectra at frequencies where the inside-to-DARM coupling functions are fully measured. Downward triangles represent the spectra at frequencies where only an upper limit for the inside-to-DARM coupling is known. Upper limits are the maximum potential value of the inside-to-DARM coupling when an injection does not generate a response in the GW strain readout. These estimates are shown against O3 background (blue line), the A+ design sensitivity (light orange line), and sensitivity estimates of Cosmic Explorer (dark orange line) and the Einstein Telescope(green line). The coupling at these planned facilities could potentially be reduced from the levels measured at LLO by further separating magnetic coupling sites, like cables and actuation magnets, from lightning-induced currents on the beam tubes.}
\label{fig:LLO_coupled_trigs}
\end{figure*}

We can estimate the effect of the individual lightning strokes on the GW strain using outside-to-inside magnetic coupling functions and inside-to-DARM coupling functions for internal magnetometers. 
We present the budget for LLO in Fig. \ref{fig:LLO_coupled_trigs}. We chose LLO to examine here due to higher nearby lightning activity relative to LHO as well as stronger observed magnetic coupling. This combination allows us to examine the worst case scenario for coupled lightning events.

To date, LIGO has not observed excess noise in GW strain due to individual lightning events \cite{schofield_150914_lightning,Davis_2021}, whereas Virgo has \cite{https://doi.org/10.48550/arxiv.2203.04014}. Note that vibrational effects from thunder have been observed both at LIGO as well as Virgo \cite{Davis_2021,https://doi.org/10.48550/arxiv.2203.04014}.
With the increasing sensitivity of the current and future GW detectors, it is important to understand if additional mitigation efforts will be required.

Similar to the budget constructed in Sec. \ref{sec:impactGWB} we use the scalar value of the outside-to-inside magnetic coupling function measured from the ratio of the amplitudes of individual lightning events as described in Appendix \ref{appendix:outsideInsideTF}. The inside-to-DARM magnetic coupling functions include both directly measured values as well as upper limits. Measured values are log-normal with a factor of 2 uncertainty, and upper limits are treated as uniform distributions from 0 to the upper limit value in each frequency bin \cite{Nguyen_2021}.

Fig. \ref{fig:LLO_coupled_trigs} includes theoretical noise curves for current and future detectors \cite{aLIGO_GWINC,Aplus_GWINC,Srivastava_2022,Hild_2011}. The predicted contribution to the DARM spectrum of the loudest, 100th loudest, and 10000th loudest individual lightning-induced magnetic events are shown (red, dark blue, and light blue, respectively) for the week analyzed (out of the 268971 total HL coincident triggers). The width of these shaded regions represents the $68\%$ confidence interval of the predicted contribution to DARM, based on the combined uncertainty in the product of the outside-to-inside and the inside-to-DARM coupling functions. The downward triangles show the high value of the $68\%$ confidence interval for values of the outside-to-inside coupling function with the upper limit inside-to-DARM coupling function, which, as an upper limit, we assume has no uncertainty.
All of these loudest events fall below the real O3 background but exceed the A+ design sensitivity below 20\,Hz. 
If, in the future, lightning would couple significantly to the GW sensitive channel at one of the detectors, the event would also be measured by multiple magnetometers at each site. This can be used to veto such events in case their predicted impact on the GW sensitive channel exceeds the sensitivity curve. This is the standard technique for PEM vetting \cite{Nguyen_2021}. One can also consider subtracting these glitches due to lightning. Glitch subtraction is performed for other types of glitches \cite{https://doi.org/10.48550/arxiv.2207.03429} and subtraction by Wiener filtering has been investigated for Schumann resonances \cite{Thrane:2013npa,Thrane:2014yza,10.1088/0264-9381/33/22/224003,PhysRevD.97.102007}.
However, the large variability and uncertainty of the magnetic coupling might make subtraction challenging.

The correlated power due to lightning could bias both unmodeled burst as well as modeled CBC searches. As an illustration,
if the magnetic coupling for third generation detectors remains the same as those for the current detectors, $\mathcal{O}(10^5)$ magnetic transients per week would appear in the data. It is expected that binary neutron star mergers could be in the sensitivity band for $\mathcal{O}(10^3s)$ in the case of third generation detectors \cite{PhysRevD.103.122004,PhysRevD.105.043010}. Modeling these coherent lightnings as a Poisson process, it is nearly guaranteed that at least one lightning transient will overlap with one of these BNS signals, with over 160 overlapping lightnings expected on average for a single event. While these coherent magnetic signals may be vetted using magnetometer measurements, their presence and rate could interfere with transient GW searches and parameter estimation. This could be mitigated to some extent by reducing the magnetic couplings, since  fewer magnetic transients coupling into the GW data stream will reduce the total impact on searches and analyses. However, we also recommend more detailed investigations into how to efficiently subtract these short transients from future longer events.

For third generation detectors, it may be important to consider facilities designs that could reduce the outside-to-inside coupling. Construction materials may need to be selected with this coupling in mind. For example, eddy current shielding from building cladding could be optimized over what is used today.

The lightning-induced beam tube currents discussed in Appendix \ref{appendix:outsideInsideTF} are probably particularly insidious because the beam tubes carry currents right into the heart of the detector where they generate magnetic fields in the most magnetically sensitive regions.  
The lightning-induced beam tube currents might be reduced by increasing the resistance of the beam tube path through the building and reducing the resistance between the beam tube and ground, or by active cancellation.  The coupling of lightning-induced beam tube currents might also be reduced by distancing the primary coupling sites, like the magnets in magnetic actuators and the cable trays that carry signal cables, from the beam tubes. Other structures may also carry induced currents, and ferromagnetic structures such as I-beams may even locally amplify the correlated magnetic fields.

Furthermore, one can try to reduce the amount of magnets and ferromagnetic components attached to the test masses, as well as moving them higher up in the seismic suspensions stages.
Finally, additional shielding can be considered as a complementary method to reduce the magnetic coupling\cite{Cirone_2019}.

\section{Conclusion}

Presented in this paper are the results of the investigation into of the effects of magnetic field fluctuations from lightning strokes and their coherence and correlations over distances of several thousands of kilometers. We discuss why they are the likely source to describe the observed coherence and correlations between magnetometers located at GW detectors LIGO Hanford, LIGO Livingston and Virgo, during their last observing run O3.

The inside-to-DARM magnetic coupling function measured at the GW detectors describes the coupling of magnetic fields to the GW sensitive strain channel for magnetic fields inside the experimental buildings. The outside-to-inside magnetic coupling function describes the additional effect to project the magnetic field inside these experimental buildings based on the magnetic field outside the building.
By using correlations between magnetic fields measured outside the buildings at the different GW detectors and using the inside-to-DARM magnetic coupling functions and the outside-to-inside coupling function, we can predict the effect of these fields on the search for a GWB. Whereas O3, was free of magnetic contamination, there is a real risk future runs might be affected by correlations from magnetic field fluctuations when LIGO and Virgo reach their enhanced sensitivities, Advanced LIGO A+ and Advanced Virgo Plus.

The effect from individual lightning strokes is projected for LLO to investigate their impact on transient GW searches. Again we found no significant effect was present during O3 but they are expected to affect the detectors when they reach design A+ sensitivity. Also future detectors such as Cosmic Explorer and the Einstein Telescope are predicted to be affected by magnetic fields from lightning strokes, assuming similar coupling functions.

To this extent we recommend improving the upper limits of the inside-to-DARM magnetic coupling function at high frequencies ( $\gtrsim 200$ Hz) by a factor of 2-3, or otherwise measuring the value in the case that it lies between the current upper limits and the ones requested. 
Furthermore more research is needed on methods to decrease the magnetic coupling, both outside-to-inside as well as inside-to-DARM. This might be crucial for third generation detectors, however current generation detectors at their final sensitivity might also benefit from reduced magnetic coupling.
The magnetic coupling could be reduced by shielding of the test masses \cite{Cirone_2019} and further optimisation of eddy current shielding from building cladding. The reduction of lightning induced beam tube currents, as discussed in Sec. \ref{sec:ImpactfutureSearches} and Appendix \ref{appendix:outsideInsideTF} should also be further investigated. Finally, for third generation detectors the placement of magnets and ferromagnetic components in the suspension design should be carefully considered.

More detailed studies are required to investigate how the effects of correlated lightning events can be mitigated at the level of the analysis. One can consider for example subtraction as is done for other noise sources \cite{https://doi.org/10.48550/arxiv.2207.03429,Thrane:2013npa,Thrane:2014yza,10.1088/0264-9381/33/22/224003,PhysRevD.97.102007}. However the large variability and uncertainty in the coupling functions might make this difficult. On the other hand, the diurnal (and seasonal) variation in lightning may be useful for distinguishing between inter-site correlations from lightning and from gravitational waves.

\acknowledgements
This material is based upon work supported by NSF's LIGO Laboratory which is a major facility fully funded by the National Science Foundation.
The authors acknowledge access to computational resources provided by the LIGO Laboratory supported by National Science Foundation Grants PHY-0757058 and PHY-0823459. 

The authors would like to thank Patrick M. Meyers and Irene Fiori for useful comments and discussions.

This paper has been given LIGO DCC number P2200241 and Virgo TDS number VIR-0800A-22.

KJ is supported by FWO-Vlaanderen via grant number 11C5720N.

MB, RF, and RMSS are supported by UO NSF grant PHY-1912604.

MWC is supported by the National Science Foundation with grant numbers PHY-2010970 and OAC-2117997.

SB acknowledges support by the NSF grant PHY-180663. 

\bibliographystyle{unsrt}
\bibliography{references}

\appendix
\section{The outside-to-inside magnetic coupling function}
\label{appendix:outsideInsideTF}

Presently, we produce estimates of the inside-to-DARM coupling of magnetic fields around the detector to the main GW strain readout by generating large fields (much larger than the magnetic fields generated by lightning) inside the buildings that house the sensitive regions of the detector, and then comparing the amplitude of the resulting signals from the GW strain readout and from the magnetometers in the buildings. In order to estimate the signal in the GW strain readout using the outdoor LEMI magnetometers, which are away from the noisy buildings and sensitive to the small inter-site correlated fields, we need to be able to estimate the amplitude of the fields inside, around the detector, from the fields measured by the outside LEMI magnetometer.

Previous estimates of this outside-to-inside magnetic field coupling were based on a comparison of outside and inside magnetometers for magnetic injections made using a generating coil that was set up at multiple locations outside of the vertex building \cite{O3Isotropic}. However, these coil-generated fields differ from the fields from distant lighting in time-scale and field uniformity. Lightning detection networks make it possible to identify lightning - generated magnetic field transients and thus to measure the outside-to-inside coupling specifically for magnetic fields from lightning.

To estimate this outside-to-inside coupling, individual lightning strokes were selected from the GLD360\footnote{For this measurement additional GLD360 data was provided by Vaisala for the week of July 8 to 14, 2019 as well as December 29, 2021 to January 3, 2022, January 25 to 27, 2022, and February 3 to 5, 2022.} at ranges exceeding 200km from LLO. The distance requirement serves to reduce any vertically-oriented field since the outdoor magnetometers are designed to detect Schumann resonance fields and have no vertical axis. We chose to calculate this for LLO due to the greater abundance of nearby lightning events. The events with the largest estimated ratio of current to distance were picked to ensure the signal was clearly witnessed by both outside and inside magnetometers. Signals were selected with an SNR exceeding 10 as measured outside and greater than 6 inside. Notch filters were used to remove all harmonics of the 60Hz mains line. All axes were added in quadrature (including the vertical components for interior magnetometers), and the result was high-passed to remove any DC offset from the quadrature sum. 

\begin{figure}
\centering
\includegraphics[width=\linewidth]{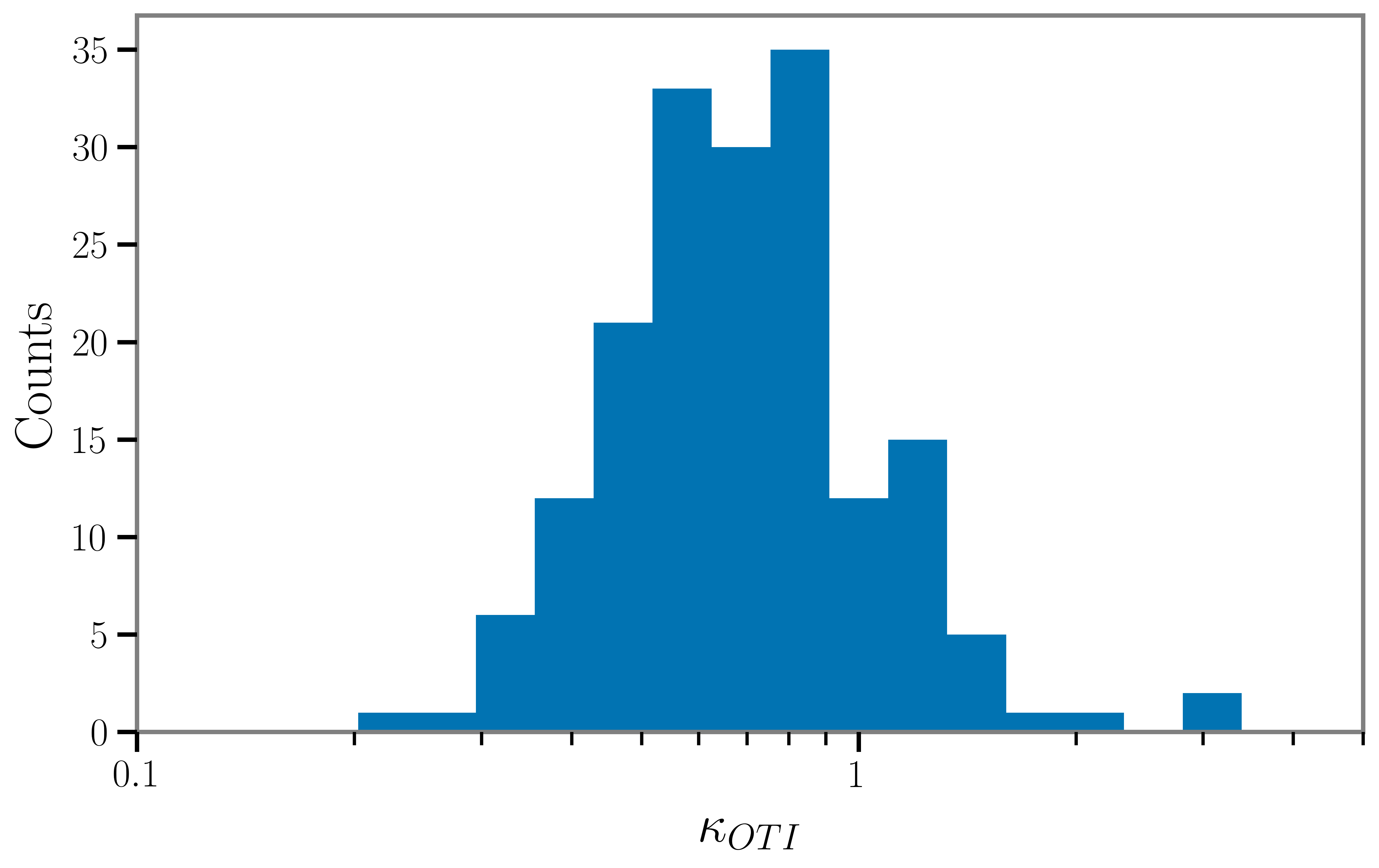}
\caption{Histogram of outside-to-inside coupling function values estimated from individual lightning strokes.}
\label{fig:outside_inside_kappa}
\end{figure}

\begin{figure}
\centering
\includegraphics[width=\linewidth]{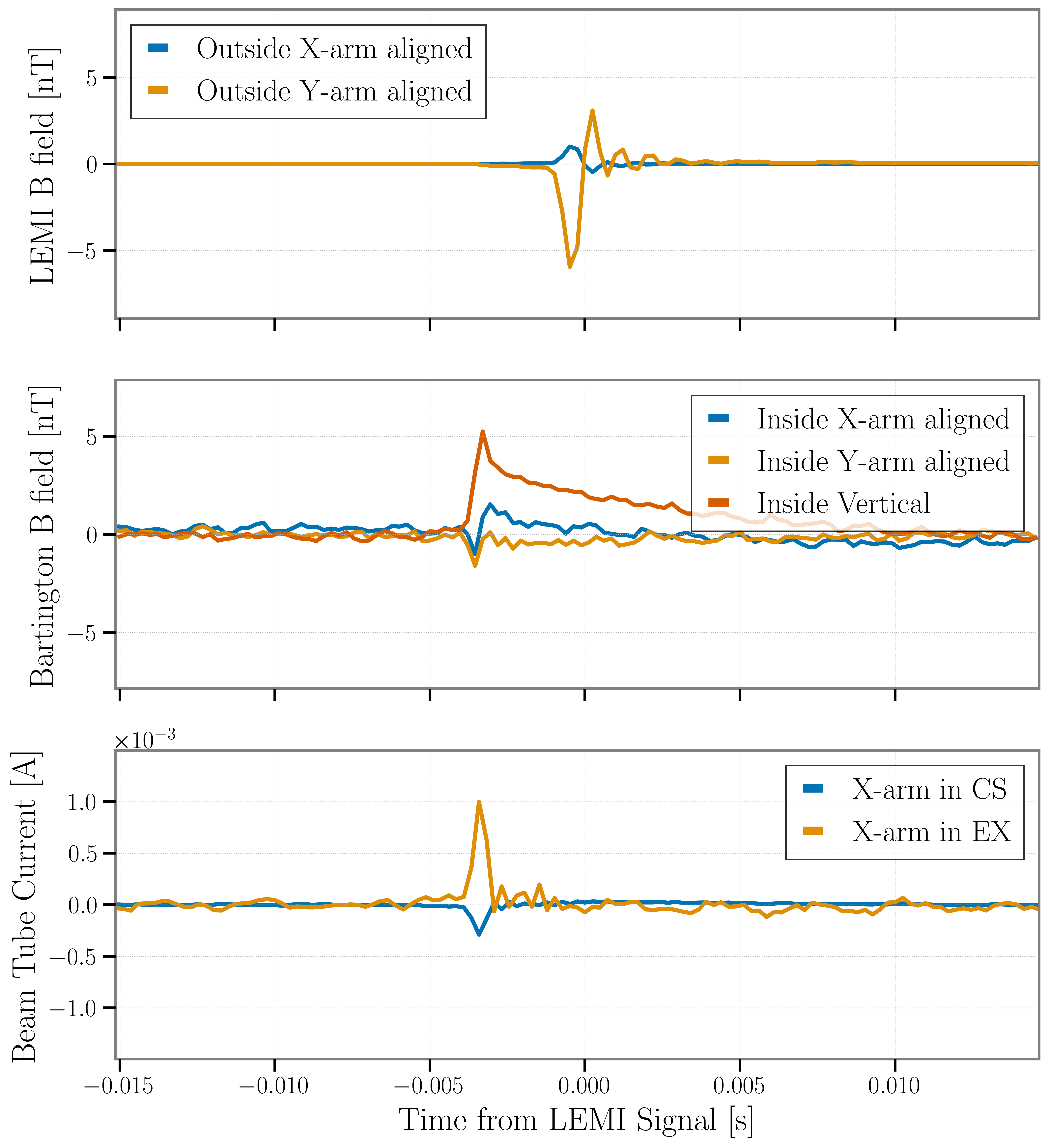}
\caption{Top: Filtered magnetometer measurement of magnetic field believed to be from a lightning stroke as measured by outdoor magnetometers at LLO. Center: The same lightning-generated field as measured by an interior magnetometer, showing a strong vertical component. Bottom: Measurements from current probes clamped to beam tube grounding cables at various locations on the X-arm (CS: corner station, EX: X end station). At the location of the magnetometer, the field from a current on the beam tube would mainly be vertical.}
\label{fig:lightning_signal_2_currents}
\end{figure}

Time series of an example lightning event are shown in the two top panels of Fig. \ref{fig:lightning_signal_2_currents}.
The ratio of the peak amplitudes of these processed time series was used to estimate a scalar value for the outside-to-inside coupling function. Using multiple lightning events results in a log-normal distribution of coupling values with mean value $0.7^{+0.4}_{-0.3}$ nT/nT (Figure \ref{fig:outside_inside_kappa}). The power-law estimate from static coil injections discussed earlier agrees strongly with this estimate, falling within the uncertainty limits for our frequency range.

Some interior magnetometers witnessed significant vertically-oriented magnetic fields for some lightning signals, primarily at LLO. Significant vertical fields are not expected for distant lightning \cite{silber2015vertical_fields} and are not monitored by the external magnetometers. Horizontal fields may produce vertical fields through induced currents or field distortions associated with ferromagnetic structures. The vertical fields that were observed in the building were roughly consistent with a simple model in which the distant lightning would induce currents in a loop consisting of the beam tube, the beam tube grounding, and the earth beneath the beam tube. This induced current would generate fields around the beam tube inside of the buildings. The existence of beam tube current transients coincident with distant lightning was confirmed using current probes clamped on beam tube grounding cables. 
GLD 360 lightning events from distances greater than 200\,km from the site were selected and the magnetic fields measured on site for these events were identified in our magnetometers in coincidence with currents measured on the beam tube.
Fig.~\ref{fig:lightning_signal_2_currents} shows one such signal witnessed by both exterior and interior magnetometers and the beam tube grounding current monitor.

\section{Error propagation for the projected impact on an isotropic SGWB}
\label{appendix:errorpropagation}

In our budget we want to combine the three different baselines by weighing them with respect to their importance for the SGWB search as introduced in Eq. \ref{eq:totMagBudget_weights}.
The the sum runs over the three baseline pairs HL, HV and LV and the definition of $\cdot C_{{\rm Mag},IJ}(f)$ was introduced in Eq. \ref{eq:C_Mag_final}.
We know that the coupling function can be divided in two components as follow,
\begin{equation}
    \Kappa_I(f) = \Kappa_{I,{\rm OTI}}(f) \cdot \Kappa_{I,{\rm ITD}}(f),
\end{equation}
where we have used the abbreviations OTI and ITD for respectively outside-to-inside and inside-to-DARM.

For the calculation of the errors we will assume all quantities are exactly known apart from the coupling functions, i.e. $s_{CSD_{IJ}} = 0$, $s_{\gamma_{IJ}} = 0$, $s_{S_{0}} = 0$ and $s_{w_{IJ}} = 0$.

The outside-to-inside coupling has only one uncertainty, whereas the inside-to-DARM has an uncertainty due to the measurement and location of witness sensors ($s_{\rm intrinsic}$) as well as due to the weekly variation ($s_{\rm weekly}$).
\begin{equation}
    s_{ \Kappa_{I,{\rm ITD}}}(f) = \sqrt{s_{\rm weekly,I}(f)^2 + s_{\rm intrinsic,I}(f)^2}
\end{equation}
The error on the total coupling function becomes
\begin{equation}
\begin{aligned}
     s_{\Kappa_I}(f) = &\left[s_{\Kappa_{I,{\rm OTI}}}^2(f) \cdot \Kappa_{I,{\rm ITD}}^2(f) + \right. \\
     & \left. s_{\Kappa_{I,{\rm ITD}}}^2(f) \cdot \Kappa_{I,{\rm OTI}}^2(f)\right]^{1/2}
\end{aligned}
\end{equation}
The error on $\hat{C}_{mag,IJ}(f)$ is given by,
\begin{equation}
\begin{aligned}
    s_{\hat{C}_{mag,IJ}}(f)= &\left[ s_{\Kappa_I}^2(f) \left(\frac{\hat{C}_{mag,IJ}(f)}{\Kappa_I(f)}\right)^2 + \right.\\
    &  \left. s_{\Kappa_J}^2(f) \left(\frac{\hat{C}_{mag,IJ}(f)}{\Kappa_J(f)}\right)^2 \right]^{1/2}
\end{aligned}
\end{equation}
Finally for the combined budget including all three baselines we get the following error
\begin{equation}
    s_{\hat{C}_{mag}}(f)  = \sqrt{\sum_{I,J}w_{IJ}^2(f) \cdot s_{C_{{\rm Mag},IJ}}^2(f)},
\end{equation}

The weekly measurements of the inside-to-DARM magnetic coupling functions at the central building of each site were used to compute the geometric standard deviation. This standard deviation is a good measure of the uncertainty introduced by the time variability of the coupling functions.
As a figure of merit we present the average geometric standard deviation for each baseline,
\begin{equation}
\begin{aligned}
    \langle s_{weekly,H} \rangle &= 1.4 \cdot \Kappa_{ H,{\rm ITD}}(f) \\
    \langle s_{weekly,L} \rangle &= 1.3 \cdot \Kappa_{ L,{\rm ITD}}(f) \\
    \langle s_{weekly,V} \rangle &= 1.6 \cdot \Kappa_{ V,{\rm ITD}}(f). \\
\end{aligned}
\end{equation}
However since there is a non negligible amount of frequency variability, the frequency dependent error will give the most accurate results.

As mentioned in the main text, the weekly error was set to  zero $s_{weekly}=0$ for the budget presented in Fig. \ref{fig:GWB_MAG_budget}. This choice was made as it is unclear whether a similar weekly variation is expected across the different coupling locations. The budget including the frequency dependent $s_{weekly}(f)$ was only marginally larger compared to the budget presented in \ref{fig:GWB_MAG_budget}.

The other errors used are:
\begin{equation}
\begin{aligned}
    s_{intrinsic,I} (f) &= 2 \cdot \Kappa_I(f) \\
    s_{OTI,+} &= 0,4\\
    s_{OTI,-} &= 0,3\\
\end{aligned}
\end{equation}

\end{document}